\documentclass[journal=apchd5,manuscript=article]{achemso}

\usepackage[version=3]{mhchem} % Formula subscripts using \ce{}
\usepackage[T1]{fontenc}       % Use modern font encodings
\usepackage[utf8]{inputenc}
\usepackage[english]{babel}
\usepackage{amsmath,amsfonts,amsthm,amssymb}
\usepackage{mathtools}
\usepackage{graphicx}
\usepackage{xcolor}
\usepackage[pdftex]{hyperref}
\usepackage{bm}
\usepackage[inline]{enumitem}
\usepackage{array}
\usepackage{gensymb}

\newcommand{\ep}{\epsilon}
\newcommand{\vp}{\varphi}

\DeclareRobustCommand{\legendsquare}[1]{%
  \textcolor{#1}{\rule{3ex}{1.5ex}}%
}
\definecolor{v}{RGB}{185,185,235}
\definecolor{b}{RGB}{180,202,235}

\author{P. A. D. Gon\c{c}alves}
\affiliation[]{Department of Photonics Engineering, Technical University of Denmark, DK-2800 Kgs. Lyngby, Denmark}
\alsoaffiliation[]{Center for Nanostructured Graphene, Technical University of Denmark, DK-2800 Kgs. Lyngby, Denmark}
\email{padgo@fotonik.dtu.dk}
\author{E. J. C. Dias}
\affiliation[]{Department of Physics and Center of Physics, University of Minho, PT-4710--057, Braga, Portugal}
\author{Sanshui Xiao}
\affiliation[]{Department of Photonics Engineering, Technical University of Denmark, DK-2800 Kgs. Lyngby, Denmark}
\alsoaffiliation[]{Center for Nanostructured Graphene, Technical University of Denmark, DK-2800 Kgs. Lyngby, Denmark}
\author{M. I. Vasilevskiy}
\affiliation[]{Department of Physics and Center of Physics, University of Minho, PT-4710--057, Braga, Portugal}
\author{N. Asger Mortensen}
\affiliation[]{Department of Photonics Engineering, Technical University of Denmark, DK-2800 Kgs. Lyngby, Denmark}
\alsoaffiliation[]{Center for Nanostructured Graphene, Technical University of Denmark, DK-2800 Kgs. Lyngby, Denmark}
\author{N. M. R. Peres}
\affiliation[]{Department of Physics and Center of Physics, University of Minho, PT-4710--057, Braga, Portugal}
\email{peres@fisica.uminho.pt}

\title[Graphene Plasmons in Triangular Wedges and Grooves]
  {Graphene Plasmons in Triangular Wedges and Grooves}

\keywords{graphene plasmons, plasmonics, nanophotonics, channel plasmons, wedge, groove}

\begin{document}

\begin{tocentry}

%\begin{figure}[h!]
%\centering
 \includegraphics{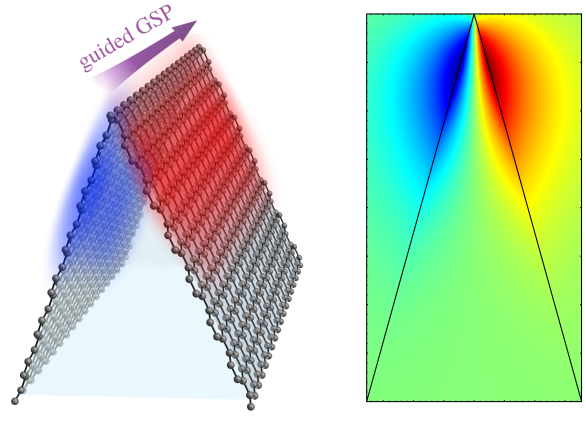}
%\end{figure}

\end{tocentry}

% *************************************************************
% ::                        ABSTRACT                     ::
% ************************************************************%%
\begin{abstract}
The ability to effectively guide electromagnetic radiation below the diffraction limit is of the utmost importance 
in the prospect of all-optical plasmonic circuitry. Here, we propose an alternative solution to conventional 
metal-based plasmonics by exploiting the deep subwavelength confinement and tunability of graphene plasmons 
guided along the apex of a graphene-covered dielectric wedge or groove. In particular, we present a quasi-analytic model 
to describe the plasmonic eigenmodes in such a system, including the complete determination of their spectrum and 
corresponding induced potential and electric field distributions. We have found that the dispersion of 
wedge/groove graphene plasmons follows the same functional dependence as their flat-graphene 
plasmons counterparts, but now scaled by a (purely) geometric factor in which all the information about the system's 
geometry is contained. We believe our results pave the way for the 
development of novel custom-tailored photonic devices for subwavelength waveguiding and localization of light based 
on recently discovered 2D materials.
\end{abstract}

\bigskip
% *************************************************************
% ::                        INTRODUCTION                     ::
% *************************************************************
%\section{Introduction}

Over the last couple of decades we have been witnessing a steady, exponential growth in the amount of information 
produced on a daily basis. In today's ``information age'', huge amounts of 
data must be processed, stored, and delivered around the world. While the data-processing part 
is still primarily carried by electronics, the routing of large volumes of information is handled with 
photonic technologies  since only these can meet the requirements in terms of high-speed, density and bandwidth. 
One of the greatest ambitions of modern nanophotonics\cite{Saleh} is to bridge the gap between electronic and photonic components, 
and ultimately to replace electronic circuits and processing units by their photonic-based counterparts.  
Current scalable photonic-based communications, however, still could not surpass the threshold towards miniaturization 
posed by the diffraction limit. In this regard, a great deal of hope\cite{Atwater} has been deposited in the sub-discipline of 
photonics known as plasmonics,\cite{MaradudinModern,Maier} which exploits the ability of surface plasmon-polaritons (SPPs)---collective oscillations of 
the free-electrons at metal/dielectric interfaces---to localize 
light into subwavelength dimensions.\cite{Nat424,difractionfree,BarnesJOA,Gramotnev:2010} Although the pursuit 
of plasmonic devices suitable for mass-production is still going on, plasmonics has already achieved some milestones, 
for instance, 
subwavelength plasmonic circuitry including waveguides, interferometers and resonators,\cite{Sci311,Sergey_nat440,Sergey_RepProgPhys,SergeyBook}  
nanolasers,\cite{PhysRevLett.90.027402,nphot6las,LPR7,Haffner:2015} 
quantum optics with or mediated by SPPs,\cite{PRL97,nnano8,nphoton8,nphysQP,Bermudez-Urena:2015} 
label-free and single molecule biochemical sensing,\cite{nl14,mrs30,nphot6,bioBook,Rodrigo10072015} 
high-resolution nanoscopy, \cite{Fang534,Liu1407} 
and even cancer theranostics.\cite{med,med2,HalasThera}

A key component in any plasmonic circuit would be an element to transfer and guide the electromagnetic (EM) fields 
from point $A$ to point $B$. Typical SPP-guiding structures\cite{PlasmWG} consist in small metallic 
stripes on a dielectric substrate,\cite{PlasmWG,MaradudinModern} 
chains of plasmonic nanoparticles,\cite{NPsWVG,NPsWVG2,guidingNPs} 
metal/dielectric/metal slabs, \cite{PlasmWG} 
or V-shaped grooves carved into a metallic substrate,\cite{AsgerRevCPP,nanoFocus,raza:2014} just to name a few. 
Among these, the latter are believed to be 
appealing candidates for subwavelength waveguiding of light since they support SPP modes 
coined as channel plasmon-polaritons (CPPs),\cite{AsgerRevCPP,MaradudinModern} which have been shown 
to deliver localized EM fields with relatively long propagation lengths,\cite{Pile:04,AsgerRevCPP,SergeyBook} 
ability to work at telecommunication wavelengths, \cite{Sergey_nat440,Moreno:06,telecom} 
and feasibility to steer the EM field along bends. \cite{Volkov:06,Sergey_nat440} 
%and demonstration of functional devices operating with CPP in triangular grooves, namely 
%waveguides, interferometers and resonators.\cite{Sergey_nat440,PlasmWG}.
% Furthermore, a theoretical work investigated the mediation of quantum entanglement between two quantum emitters 
%placed within the V-groove.\cite{PhysRevLett.106.020501}
%
The earliest reference on CPPs can be found in a theoretical investigation carried out by Dobrzynski and Maradudin, 
having obtained analytic expressions in the electrostatic limit for an infinitely sharp wedge.\cite{PhysRevB.6.3810} 
Many subsequent works then followed 
in similar wedge 
configurations with a rounded edge.\cite{PhysRevB.24.5703,PhysRevB.32.6045,PhysRevB.66.035403,PhysRevB.42.11159} 
During the past decade---owing to the rapid progresses in 
nanofabrication and computational tools---a renewed interest has emerged on 
CPPs guided along triangular grooves sculpted in metal substrates, leading to a plethora of 
theoretical and experimental studies.\cite{AsgerRevCPP,nl9_nanoFocus,Moreno:06,Bozhevolnyi_oe17,Bozhevolnyi:09,nl10:Luo,raza:2014}.

In recent years, graphene\cite{Geim09,RMP81}---an atomically-thin $sp^2$-hybridized carbon allotrope 
in which the atoms sit at the vertexes of a honeycomb lattice---has come to the light as a novel 
plasmonic material.\cite{GoncalvesPeres,AbajoACSP,Xiao2016,Primer,ACSgp} Graphene is classified as 
a two-dimensional (2D) semi-metal whose charge-carriers 
exhibit a linear dispersion.\cite{Geim09,RMP81} When doped, graphene also sustains plasmon-polaritons that inherit the 
extraordinary optoelectronic properties of this material. In particular, gate-tunable graphene surface plasmons (GSPs) 
have been shown to deliver highly confined EM fields into deep subwavelength regions, large field-enhancements, 
strong light-matter interactions, and carry the prospect of low-loss plasmonics.\cite{GoncalvesPeres,AbajoACSP,Xiao2016,Primer,ACSgp,nlgp} 
In addition, the ability to easily control the carrier-density in graphene, e.g. by electrostatic gating and/or chemical means, 
constitutes a major advantage of GSPs over conventional metal-based plasmonics. 
Popular configurations to realize GSPs involve the nanostructuring of an otherwise continuous graphene sheet into 
graphene ribbons,\cite{NatNano,nphoton7,luxmoore14,Rodrigo10072015,natcomm_Gribbons} 
disks,\cite{NJP14,ACS7,ZFnl14,nl14XZ,AsgerTinyDisks} rings,\cite{ACS7,NJP14} and 
graphene anti-dots (either as individual structures or in periodic arrays).\cite{nl14XZ,Optica2,PlasmonicCrystals}

Here, we propose a different approach to deliver strongly localized GSPs 
which does not involve any nanopatterning done on the graphene layer; it simply consists in depositing 
graphene onto a V-shaped wedge or groove previously sculpted in the receiving substrate (a different, 
but related configuration was the subject of a previous numerical study\cite{Liu:13}). This can be done by employing the same 
techniques used to fabricate metallic grooves,\cite{AsgerRevCPP} followed by the graphene deposition 
or even direct-growth on a pre-configured copper substrate.\cite{PhysRevB.83.245433} Other possibilities include 
folding a graphene layer or by exploring the formation of wrinkles 
(either naturally occuring\cite{V-graphene,C6NR01992G,wrinklesElectronic} 
or deliberately formed\cite{PhysRevB.83.245433}). In this way, 
one departures from customary flat-graphene geometries and effectively produces a 1D channel which not only confines 
light in the vertical direction that bisects the channel, but is also capable of producing lateral confinement of EM radiation. 

\begin{figure}[h]
  \centering
    \includegraphics[width=0.4\textwidth,clip]{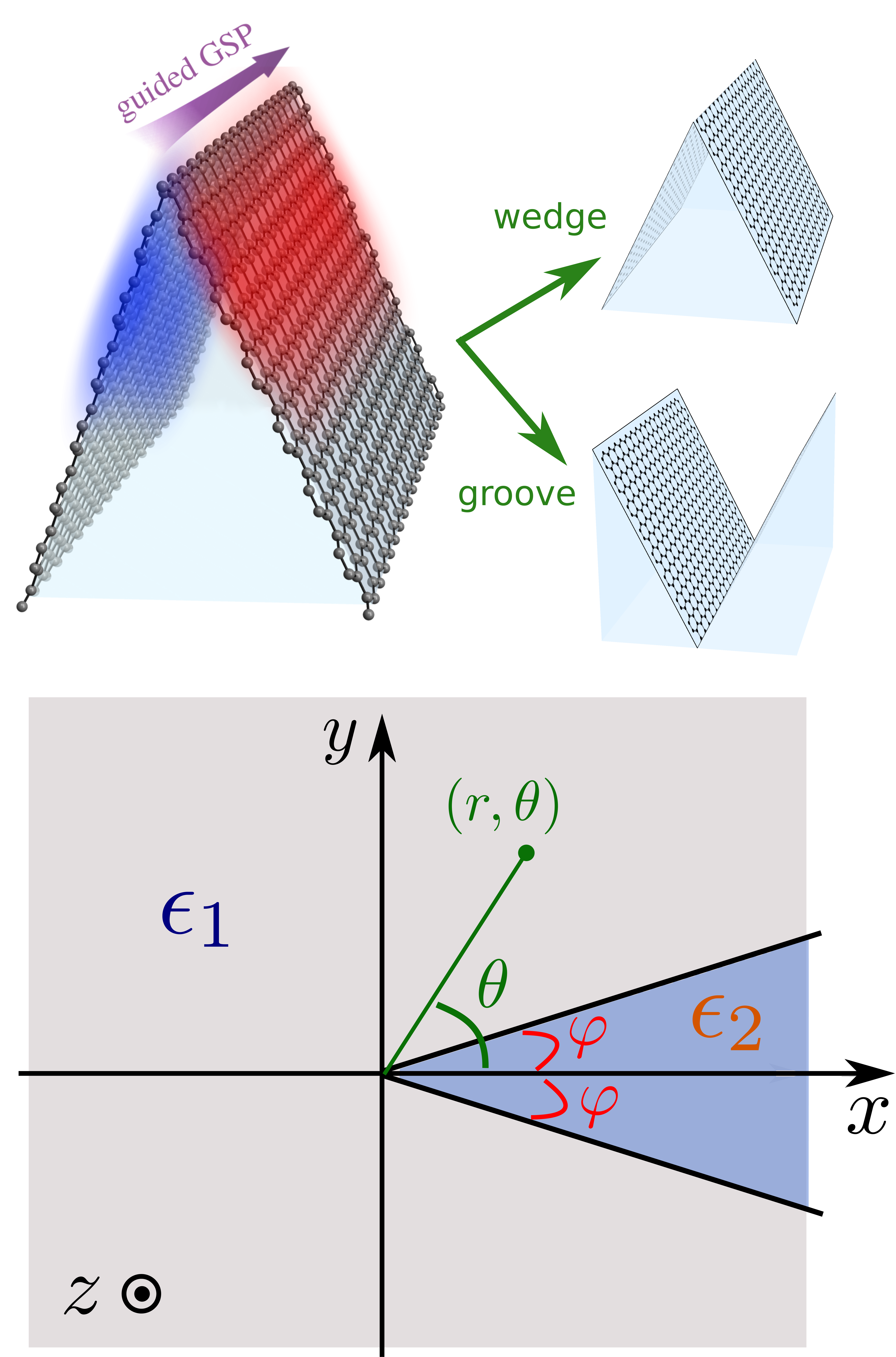}
      \caption{Upper panel: pictorial representation of a guided GSP mode along a dielectric wedge; we denote the system a
      \emph{wedge} (\emph{groove}) whenever the medium filling the $2\vp$-open region possesses a higher (lower) 
      value for the relative permittivity. Lower panel: coordinate system for a wedge or groove interface---with an opening angle of $2\varphi$---
      between two dielectric media characterized by relative permittivities $\ep_2$ (for $-\vp \leq \theta \leq \vp$) 
      and $\ep_1$ (for $\vp \leq \theta \leq 2\pi-\vp$).}\label{fig:SYS}
\end{figure}

In this work, we present a quasi-analytic method to derive the dispersion relation and corresponding spatial distributions 
of the potential and electric fields akin to GSPs guided along a V-shaped channel. 
We shall consider both the wedge and groove geometries---see Fig.~ \ref{fig:SYS}.
In what follows we work within the electrostatic limit, which turns out to be a very good approximation for GSPs owing to the 
large wavevectors carried by plasmons in graphene (and thus retardation becomes unimportant).
Interestingly, we find that for a fixed wedge/groove angle the 
corresponding wedge/groove graphene plasmon (WGP/GGP) dispersions follow a universal scaling law 
that depends purely on the system's geometry. Thus, 
by performing the calculations for a given angle, $\varphi$, one immediately gains complete knowledge of the 
WGPs/GGPs' wavevectors for every frequency---in other words, all the modes and corresponding dispersion 
relations are obtained at once. This constitutes an enormous advantage in terms of computational resources and time when 
gauged against full-wave numerical simulations. From a device-engineering perspective this scaling property 
should also significantly ease design of waveguides for given applications. Finally, 
in further support of the accuracy of our quasi-analytic technique, we 
have also performed rigorous electrodynamic simulations with the aid of a commercially available (Comsol MultiPhysics) finite-element method (FEM), 
to which we have obtained a remarkable agreement.

We demonstrate that by using graphene-covered triangular wedges or grooves one can harness the unique properties 
of GSPs to create novel 1D subwavelength plasmonic waveguides 
that can squeeze light into deep subwavelength regimes. This becomes particularly relevant at THz and mid-IR 
frequencies since traditional metal-based plasmonics perform poorly in this spectral range (resembling freely 
propagating light).\cite{MaradudinModern} Furthermore, we believe that this work can set the stage for future investigations of 
graphene plasmons in 1D channels, with potentially relevant applications ranging from plasmonic circuitry 
and waveguiding to biochemical sensing with WGPs/GGPs or their integration with microfluidics on a chip.

% ================================================== // ==================================================

%\section{Theory}
\section*{Results and Discussion}

% *************************************************************
% ::                 THEORY AND RESULTS                      ::
% *************************************************************

We consider an idealized geometry in which a graphene monolayer is sandwiched between a triangular dielectric wedge (or groove) with relative 
permittivity $\ep_2$ and a capping dielectric material with relative permittivity $\ep_1$, as depicted in Fig. \ref{fig:SYS}. 
As it will 
become apparent later, our model is completely general irrespective of the specific values for the dielectric 
constants of the cladding insulators. However, in order to cope with the standard nomenclature, 
we shall refer to a \emph{wedge} whenever $\ep_2 > \ep_1$ and vice-versa to denote a \emph{groove}. 
In addition, albeit here we are primarily interested in graphene, the theory outlined below can be applied to any 2D 
layer deposited onto the triangular shape, be it a 2D electron gas or a doped 2D transition metal dichalcogenide (TMDC), etc.

Before proceeding to the description of our quasi-analytic method, 
we first bring to the reader's attention that one can treat the cases of even and odd symmetry in the potential (or induced charges) 
--- with respect to the line bisecting the triangular cross-section---separately, 
as this makes the problem more amenable to handle. In particular, for the case of even symmetry (i.e., when the induced 
charges are symmetric in the graphene half-planes which constitute the V-shape), we have found that these even-symmetry modes 
are not highly confined near the apex of the wedge/groove, with their dispersion being virtually the same as for GSPs in a flat, 
planar dielectric/graphene/dielectric interface (cf. SI). 
Conversely, as it will become clear ahead, the corresponding 
odd WGPs/GGPs modes exhibit strong field confinement near the apex of the wedge/groove, 
and therefore we shall limit our discussion solely to the odd-symmetry case hereafter. 
Owing to the high localization of the field near the apex, we note that although we assume (for simplicity) 
an infinitely long V-shape, the theory developed here remains adequate in the description of V-structures of 
finite height/depth as long as their size is larger than the region spanned by the field distribution along the axis of symmetry.

The extremely large wavevectors (when compared with light's free-space wavevector, $k_0=\omega/c$) attained by graphene plasmons 
allow us to treat plasmonic excitations in graphene within the electrostatic limit with high accuracy. In this regime, the 
induced electric potential akin to GSPs must satisfy Poisson's equation, which in cylindric coordinates reads 
\begin{equation}
 \left[ \frac{\partial^2}{\partial r^2} + \frac{1}{r} \frac{\partial}{\partial r} 
 + \frac{1}{\partial r^2} \frac{\partial^2}{\partial \theta^2} - q^2 \right] \Phi(r,\theta) = - \frac{\rho(r,\theta)}{\ep_0}\ ,
\label{def:Poisson_2D}
\end{equation}
where we have written the scalar potential as $\Phi(\mathbf{r})  = \Phi(r,\theta) e^{iqz}$, 
owing to the system's translational invariance along the $z$-axis (an implicit time-dependence 
of the form $e^{-i\omega t}$ is assumed). 
This effectively reduces our initial 3D problem into a 2D one, and will allow us to parameterize the dispersion relation 
of the guided GSPs in terms of the propagation constant $q$, i.e. $\omega \equiv \omega(q)$. Hence, 
the solution of Eq.~(\ref{def:Poisson_2D}) renders the WGPs/GGPs modes which propagate along the longitudinal direction. 
Formally, the solution of this equation in the medium $j=1,2$ can be written as
\begin{equation}
 \Phi (r,\theta) = 
 \frac{i\sigma(\omega)}{\omega} \int_{0}^{\infty} dr' G_j (r,\theta;r',\vp) 
 \left[ \frac{\partial^2}{\partial r'^2} - q^2 \right] \Phi(r',\vp) \ ,
 \label{eq:integro_diff_phi_wedge}
\end{equation}
where $\sigma(\omega)$ is the dynamical conductivity of graphene,
and $G_j (r,\theta;r',\vp)$ is the Green's function associated with Eq. (\ref{def:Poisson_2D}) in that medium; 
the latter is defined explicitly in the SI. Moreover, when writing the preceding equation, 
we have expressed the carrier-density as $\rho(r,\theta) = -e n(r) \delta(\theta-\vp) / r$, where the 2D 
particle density, $n(r)$, was written in terms of the electrostatic potential 
by combining the continuity equation together with Ohm's law (cf. SI). We further remark that 
we only need to solve for the potential in, say, the upper-half space ($0 \leq \theta \leq \pi$), 
since we are looking for solutions in which the potential is odd with respect to the symmetry axis.
It is clear from Eq. (\ref{eq:integro_diff_phi_wedge}) that the potential in the \emph{whole space} 
can only be derived once the potential \emph{at the graphene sheet} (i.e., $\theta=\vp$) is determined.
To that end, we set $\theta=\vp$ and then employ an orthogonal 
polynomials expansion technique\cite{WU1986795,PRB84.085423,GoncalvesPeres} to transform the above integro-differential 
equation for the potential at the graphene, $\phi(r) \equiv \Phi(r,\vp)$, into a standard linear algebra eigenproblem. This is 
done by expanding the electrostatic potential evaluated at the graphene layer as 
$
%\begin{equation}
 \phi(r) = \sum_n c_n L_n^{(0)}(qr) e^{-qr/2} %\ , \label{eq:expansion_Laguerre_poly}
%\end{equation}
$
, with $L_n^{(0)}$ denoting the generalized Laguerre polynomials,\cite{AS} and where the $c_n$'s are the entries of the 
eigenvectors defined by the following eigensystem (obtained by exploiting the appropriate orthogonality relations\cite{AS}):
\begin{equation}
 \frac{i\omega}{q\sigma(\omega)} c_m = \sum_{n=0}^{\infty} U_{mn} c_n \ , \label{matrixEq:wedge}
\end{equation}
where the matrix elements $U_{mn}$ read 
\begin{align}
 U_{mn} = \int_{0}^{\infty} \int_{0}^{\infty} dx dy\ G (x,\vp;y,\vp) e^{-\frac{x+y}{2}} L_m^{(0)}(x) \nonumber\\
 \times \left[\frac{3}{4} L_n^{(0)}(y) - L_{n-2}^{(2)}(y) - L_{n-1}^{(1)}(y) \right] \ . \label{Umn:wedge}
\end{align}
We note that the double integration over the dimensionless variables $x=qr$ and $y=qr'$ can be performed analytically, thereby making the 
computation of the matrix elements extremely fast. Notice that we have dropped the index $j$ in the Green's function because the boundary 
condition at $\theta=\vp$ enforces that $G_1 (x,\vp;y,\vp) = G_2 (x,\vp;y,\vp)$, and therefore one can choose either Green's function arbitrarily 
without any loss of generality.

The eigenvalue equation (\ref{matrixEq:wedge}) can be solved numerically using standard linear-algebra routines. 
Once we find the corresponding 
eigenvalues $\tilde{\lambda}_n$ (whose number matches the length of the vector $\vec{c}$, i.e. 
$N+1$, where $N$ truncates the expansion for $\phi(r)$, and convergence was checked empirically---cf. SI), the spectrum 
of graphene plasmons traveling along the triangular wedge/groove straightforwardly follows from
\begin{equation}
 \frac{i\omega}{q\sigma(\omega)}  = \tilde{\lambda}_n \ , \label{lambda:wedge}
\end{equation}
where, for a given opening angle $2\vp$, Eq. (\ref{lambda:wedge}) returns a discrete set of WGPs/GGPs modes. We stress that 
all the momentum and frequency dependence stems from the LHS of the previous equation; hence, 
the eigenvalues $\tilde{\lambda}_n \equiv \tilde{\lambda}_n(\vp)$ carry a purely geometric meaning since 
they depend uniquely on the configuration of the system (opening angle and material parameters). In particular, using graphene's 
Drude-like conductivity with negligible damping,\cite{GoncalvesPeres} one obtains a ``universal scaling law'' for the 
dispersion relation of wedge/groove graphene plasmons,
\begin{equation}
 \Omega(q)  = \Omega_{\mathrm{flat}}(q) \frac{2}{\pi} \sqrt{ \lambda_n } \ , \label{spectrum:wedge}
\end{equation}
where the relation $\tilde{\lambda}_n = \frac{4}{\pi^2 \ep_0 (\ep_1 + \ep_2)} \lambda_n$ has been used, 
and stems from factorizing a constant proportionality factor entering in the Green's function (see the text after Eq. (S33) in the SI). 
The above equation gives the energy of the guided graphene plasmon modes parameterized by the propagation constant along the apex of the wedge. Here, 
$\Omega_{\mathrm{flat}}(q) = \sqrt{\frac{4 \alpha \hbar c}{\ep_1 + \ep_2} E_F q}$ 
is simply the dispersion relation followed by GSPs in flat graphene\cite{GoncalvesPeres} sandwiched between two dielectrics with $\ep_1$ and $\ep_2$ 
(where $\alpha\simeq 1/137$ denotes the fine-structure constant).
Notice that once we have determined $\lambda_n$, we possess complete knowledge of the 
WGPs/GGPs spectrum---for any point in the entire $(q,\omega)$-space ---, 
all of this with only \emph{one} computation. In fact, we can even plot the dispersion of distinct 2D materials that support SPPs modes 
from a single computation of $\lambda_n$, since the latter does not depend on the 2D conductivity that characterizes 
the particular 2D material [recall Eq. (\ref{lambda:wedge})].
It is instructive to note that, in general, the spectrum of WGPs/GGPs contains a discrete set of even and odd modes 
(although here we describe only modes with odd-symmetry for the reason stated above in the text), in a similar way 
to SPPs supported at metallic wedges/grooves.\cite{AsgerRevCPP} This is a consequence of the lateral confinement 
near the tip of the wedge (or the bottom of the groove), 
and bears some resemblance to finding the electronic eigenstates of a particle in 
a quantum wire.\cite{harrison}

\begin{figure*}%[h]
  \centering
    \includegraphics[width=0.6\textwidth]{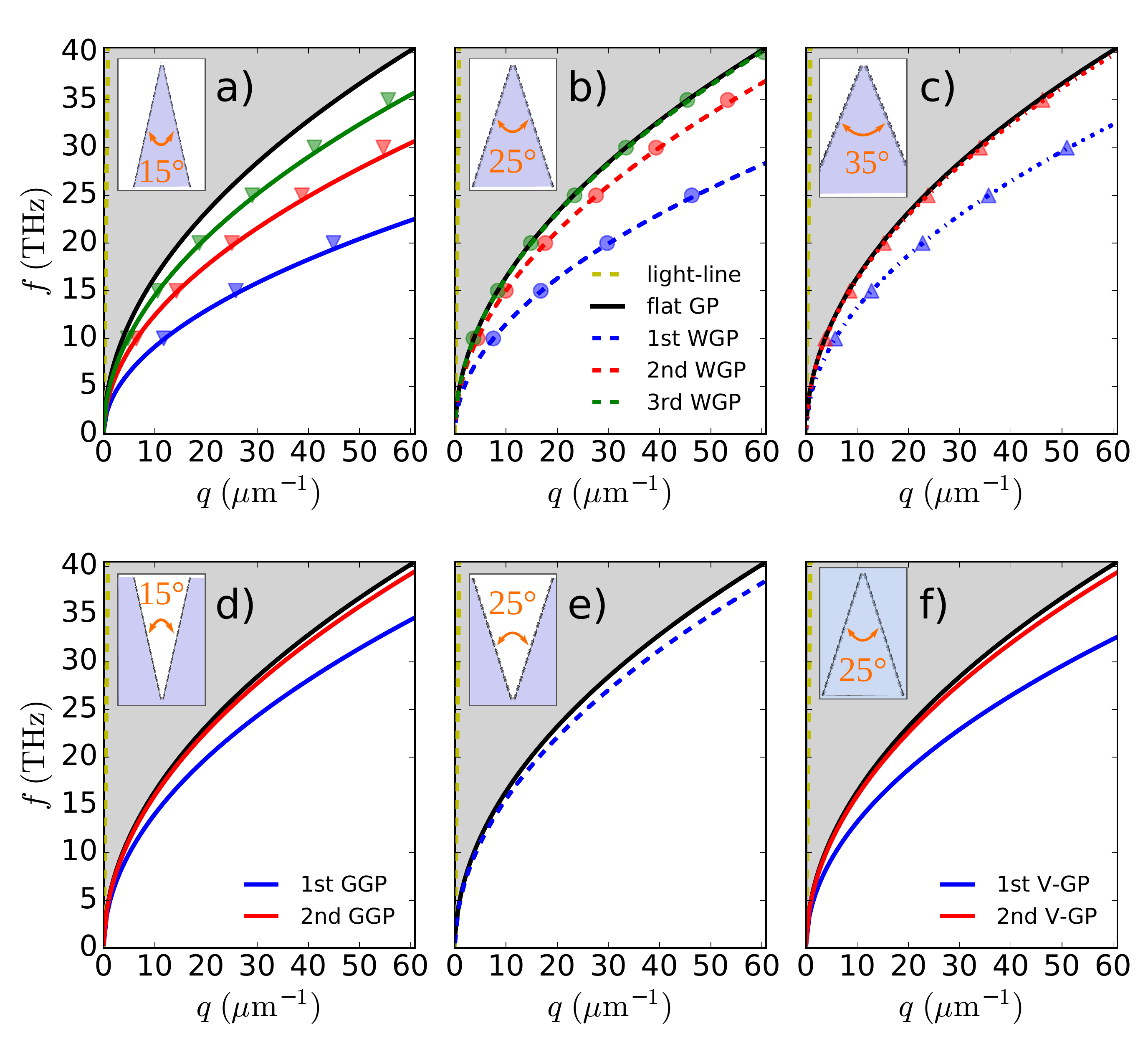}
      \caption{Spectrum of WGPs (upper row) and GGPs (first two bottom panels) sustained in different graphene-covered triangular configurations, 
      with different opening angles, $2\vp$ (indicated in the insets), as given by Eq. \ref{spectrum:wedge} [we take $E_F=0.4$ eV]. 
      The solid black line represents the dispersion 
      of GSPs in a flat interface and serves as reference. The colored $\bigtriangledown$, $\bigcirc$ and $\bigtriangleup$ data points 
      figuring in the upper row correspond to the results for the WGP dispersion as obtained from full-wave numerical simulations 
      (COMSOL's finite-element method). The insets' shading \legendsquare{v} represents an insulator with $\ep=4$, whereas 
      the white regions denote a medium with $\ep=1$ (e.g. air). The last panel shows the spectrum of GSPs guided along a 
      V-shaped graphene embedded in a homogeneous medium with $\ep=2.5$ (shaded in the inset as \legendsquare{b}).
      }\label{fig:spectrum}
\end{figure*}

\begin{figure}[h]
  \centering
    \includegraphics[width=0.4\textwidth]{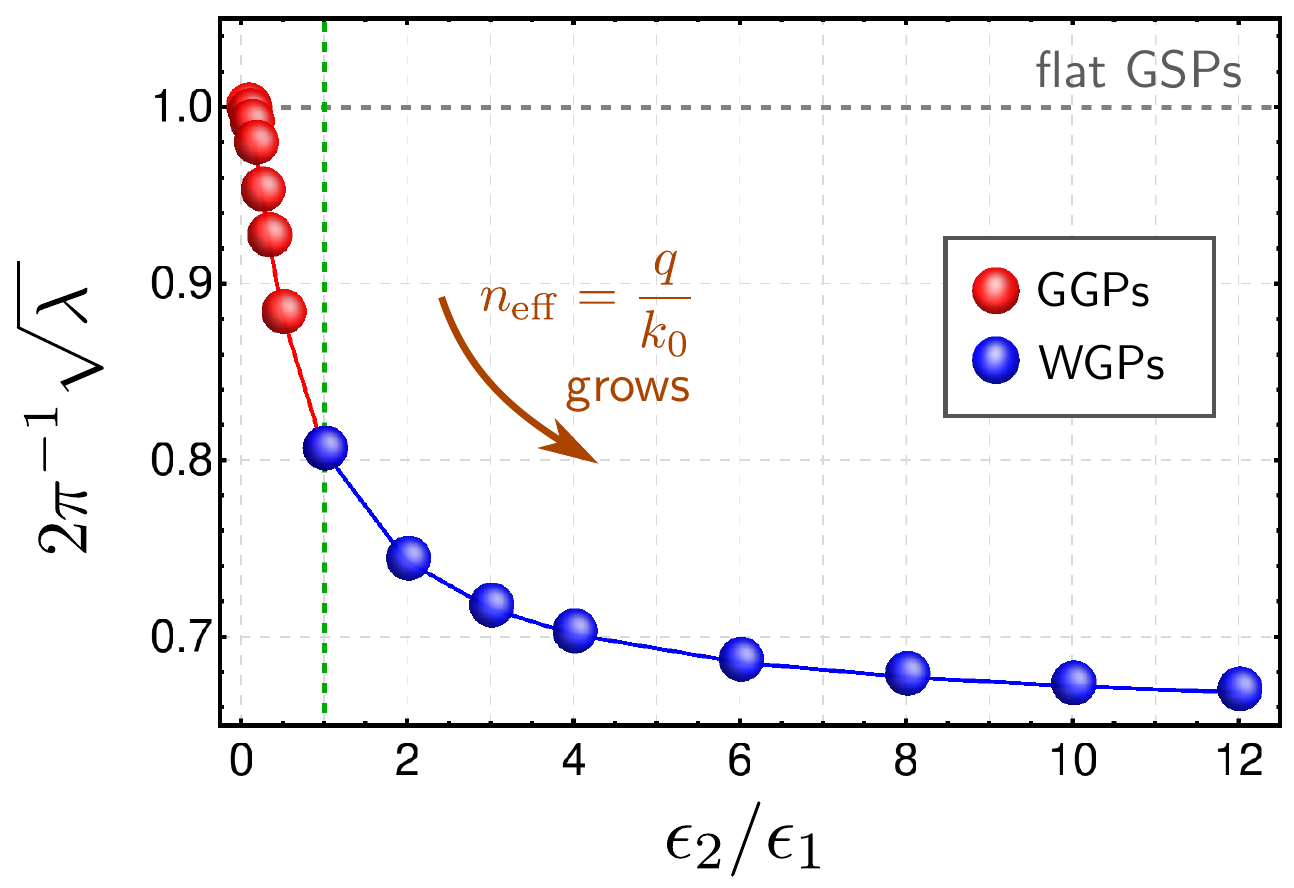}
      \caption{Dependence of the eigenvalues akin to the fundamental mode---in fact, 
      of the proportionality factor, $2\pi^{-1} \sqrt{ \lambda}$, 
      figuring in Eq. (\ref{spectrum:wedge})---as a function of the ratio $\epsilon_2/\epsilon_1$. In computing the 
      data in blue we have fixed $\epsilon_1=1$ while varying $\epsilon_2$, and vice-versa for the red data points. 
      The vertical green dashed line indicates the point where $\epsilon_1=\epsilon_2=1$, whereas the horizontal gray dashed line 
      sets the upper bound corresponding to the flat GSP dispersion. We further note that 
      the value of $\lambda$ is also sensitive to the absolute value of the dielectric constant that is kept constant while 
      the other varies.      
      }\label{fig:EpsRatio}
\end{figure}

In Fig. \ref{fig:spectrum} we have plotted the dispersion relation of graphene plasmons guided along 
the edge of triangular wedges and grooves with different opening angles, $2\vp$ (indicated in the insets), which, as we have 
already anticipated, consists in a discrete set of well-defined modes with increasing energy. The figure 
plainly shows that the spectrum of both WGPs and GGPs strongly depend on the angles of the triangular opening, with 
smaller angles rendering correspondingly larger plasmon wavevectors for the same frequency, which in turn is 
an indication of stronger field confinement near the apex of the V-shape. 
Another striking feature visible in the figure is the outstanding agreement between the quasi-analytic theory detailed 
above and the electrodynamic simulations using the FEM technique (see SI 
for details on the simulations). This constitutes further evidence of 
the ability of our quasi-analytic method to render accurate results, 
while also providing a deeper fundamental understanding of the scaling properties.
Also from the inspection of Fig.~\ref{fig:spectrum}, a strong contrast can be perceived 
between the dispersion curves akin to WGPs and GGPs, demonstrating the 
superiority of the former in squeezing light below the diffraction limit as they attain larger wavevectors for 
the same angle of the structure. We further note that one can transform a wedge into a groove and vice-versa either by swapping the values of 
$\ep$ or by applying the angular transformation $\vp \rightarrow \pi-\vp$ (this essentially interchanges 
the Green's functions $G_1$ and $G_2$). For the sake of completeness, in Fig. \ref{fig:spectrum}f panel we have portrayed the 
spectrum of GSPs guided along a triangular apex embedded in a homogeneous dielectric medium with the same average 
relative permittivity as its wedge and groove counterparts. It can be observed that---for the same angle---each 
of the modes attain increasingly larger wavevectors as we move from a groove, embedded and wedge configuration (in this order). 
This hints us that the ability to reach deep subwavelength regimes strongly depends on the $\ep_2/\ep_1$ ratio, for a 
fixed (acute) angle. Such prediction is confirmed by Fig. \ref{fig:EpsRatio}, 
in which we observe that the scaling factor appearing in Eq. (\ref{spectrum:wedge}) decreases monotonically 
with increasing $\ep_2/\ep_1$. In turn, this translates into higher effective indexes, $n_{\mathrm{eff}} = q/k_0$, 
for larger quotients $\ep_2/\ep_1$. As an example, we have obtained $n_{\mathrm{eff}} \simeq 72$ for a frequency of 
$20$ THz in a configuration corresponding to panel in Fig.~\ref{fig:spectrum}b. Even larger 
effective indexes can be obtained at higher frequencies (for instance, for the CO$_2$ laser wavelength of $10.6\, \mu$m this 
value climbs to $n_{\mathrm{eff}} \simeq 100$, using the same parameters).
Naturally, the higher amount of field localization promoted by the WGP/GPP modes comes hand in hand with slightly larger propagation losses (cf. SI), 
a trait that is well-known and characteristic of plasmonics. Still, we have found that the number of plasmon oscillations within a 
propagation length \cite{Maier,GoncalvesPeres} remains unaltered when comparing between different mode orders or flat-GSPs (see SI for further details). Indeed, owing to the ``universal scaling'' 
epitomized by Eqs. (\ref{lambda:wedge}) and (\ref{spectrum:wedge}), the \emph{ratio} between loss and field confinement is independent 
of the mode order and is the same as for flat-GSPs.

Before concluding the analysis of Fig.~\ref{fig:spectrum}, we highlight the capability of these guided GSPs modes to render 
extreme light-localization. This ability can be appreciated by considering the distance of their dispersion curves to the 
light-line (yellowish dashed-line 
near the vertical axis) and the large effective indexes that were obtained. 
This departure from the light-cone also justifies the high accuracy in the treatment of these modes 
within the electrostatic limit, as retardation effects are negligible and GSPs possess an essentially electrostatic character.\hfill\\

\begin{figure}[h]
%\vspace*{1mm}
  \centering
    \includegraphics[width=0.45\textwidth]{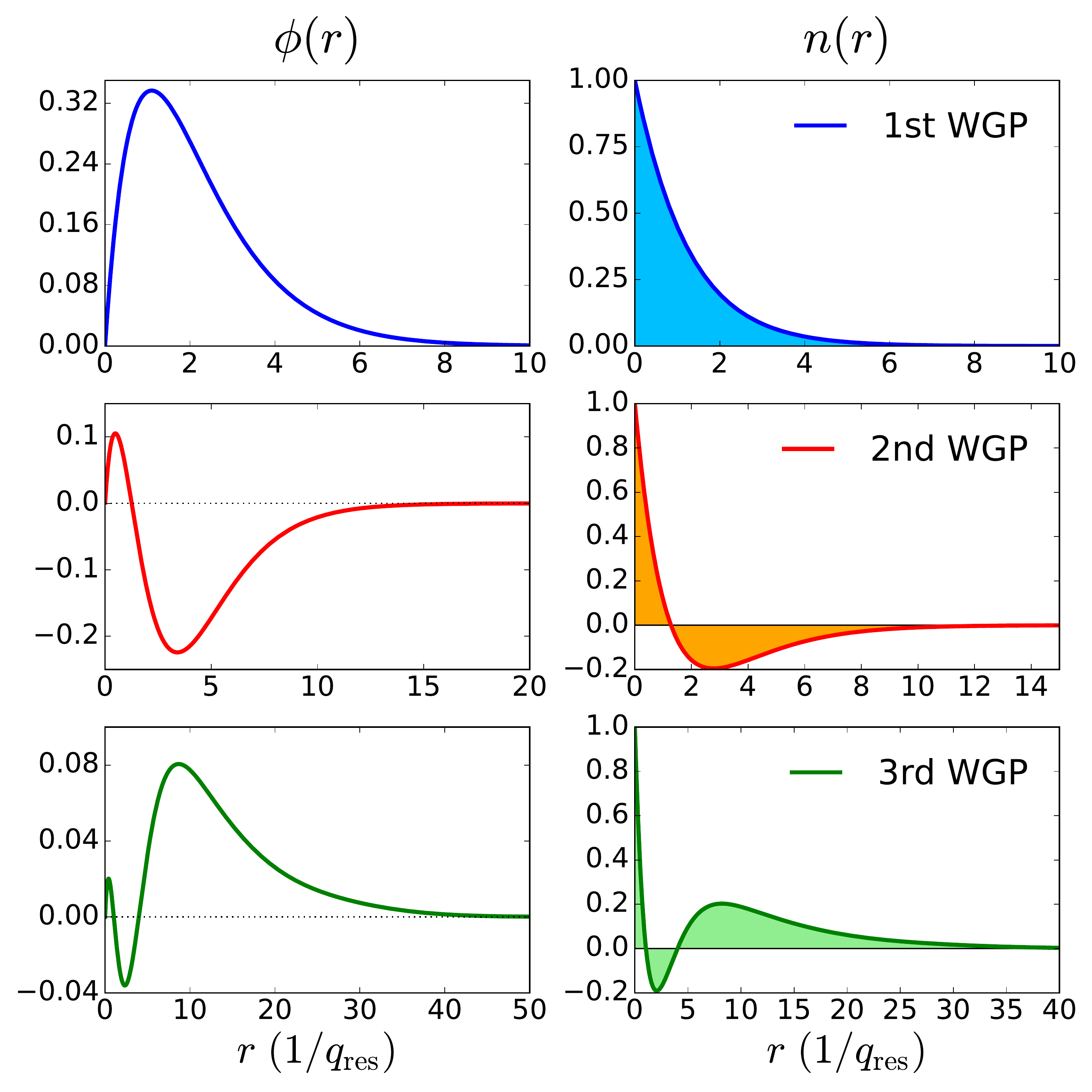}
      \caption{Radial distributions of the electrostatic potential, $\phi(r) \equiv \Phi(r,\vp)$, and 2D particle density, $n(r)$, 
      for the three lowest eigenmodes of a $25\degree$ wedge covered with graphene (in a.u.), at a resonant frequency 
      of $f=20$ THz. The radial coordinate is plotted in units of the inverse wavevector akin to each mode [obtained by inverting 
      Eq. (\ref{spectrum:wedge}) for a angular frequency of $\omega = 2\pi \times 20$ THz].
      The remaining parameters are the same as in Fig. \ref{fig:spectrum}b (i.e., $\ep_1=1$, $\ep_2=4$, and $E_F=0.4$ eV).
      }\label{fig:in_plane}
\end{figure}

\noindent
{\bf Potential and electric-field distributions.} Furthermore,
we remark that the solution of the eigenproblem in Eq.~(\ref{matrixEq:wedge}) also allows us to reconstruct the
electrostatic potential within the graphene by feeding the obtained eigenvalues and eigenvectors back into the expansion for $\phi(r)$. From here, 
the 2D particle-density 
directly follows via $n(r) = \frac{\sigma(\omega)}{i e \omega} \left[ \partial^2 / \partial r^2 - q^2 \right] \phi(r)$ (see SI). 
Both these physical quantities, evaluated at the graphene's surface, that is $\theta=\vp$ 
(at $\theta=-\vp$ the distributions are antisymmetric), are shown in Fig.~\ref{fig:in_plane}. Here, we consider WGPs 
with frequency $f=20$ THz guided along a $2\vp = 25\degree$ triangular wedge (which corresponds to the case illustrated 
in Fig. \ref{fig:spectrum}b).

Figure \ref{fig:in_plane} indicates that the highest density of charge-carriers occurs at the apex of the wedge, irrespective of the mode order, 
although the density drops towards zero at increasingly larger distances from the origin for the higher-order modes. In addition, 
we note that $q_\mathrm{3rd}^{-1} > q_\mathrm{2nd}^{-1} > q_\mathrm{1st}^{-1}$ for the same frequency, which makes 
the difference in the confinement even more dramatic. We emphasize that 
the number of nodes of both $\phi(r)$ and $n(r)$ is given by $m-1$, where $m=1,2,...$ (for modes 
below the flat GSP dispersion curve) stands for the mode order. On the other hand, the value of the 
potential evaluated at the graphene layer is large \emph{near} the vertex of the triangular cross-section, specially for the 
fundamental mode (where it is maximum). We acknowledge that, however, it is not located \emph{exactly} at $r=0$. 
This is consequence of the infinitely sharp apex, whose corresponding Green's functions 
strongly oscillate at very small $r$, since they are not well-defined (but are bounded, i.e. do not diverge)
in the $r \rightarrow 0$ limit.
Conversely, for modes with higher energy, the potential tends to shift its weight farther 
from the apex of the V-structure as the mode order increases. Such behavior was already expected in the light of Fig.~\ref{fig:spectrum}, 
since the WGP wavelength becomes smaller as the mode order decreases; as a consequence, 
the fundamental WGP mode exhibits the highest field-confinement, thereby being able to probe deeper 
into the V-wedge owing to its shorter wavelength when compared to the higher branches of the polaritonic spectrum. 
Again, we stress that although our model assumes infinitely deep triangular cross-sections, it can accurately 
describe finite-sized V-shapes as long as the height (depth) of the wedge (groove) is located at a distance somewhere 
along the ``tail'' of the quantities plotted in Fig. \ref{fig:in_plane} (where they are essentially zero).
In passing, we note that apart from the plasmon modes located at the apex of, say, a wedge ($2\vp$ opening angle), the modes 
sustained at the corresponding ``grooves'', originating from the truncation of the structure---forming a $\pi/2+\vp$ angle---, can also be determined
using the same guidelines as above, provided that the height of the wedge is large enough to prevent the hybridization of the modes. The same 
reasoning also holds for grooves.

\begin{figure}[h!]
  \centering
    \includegraphics[width=0.475\textwidth]{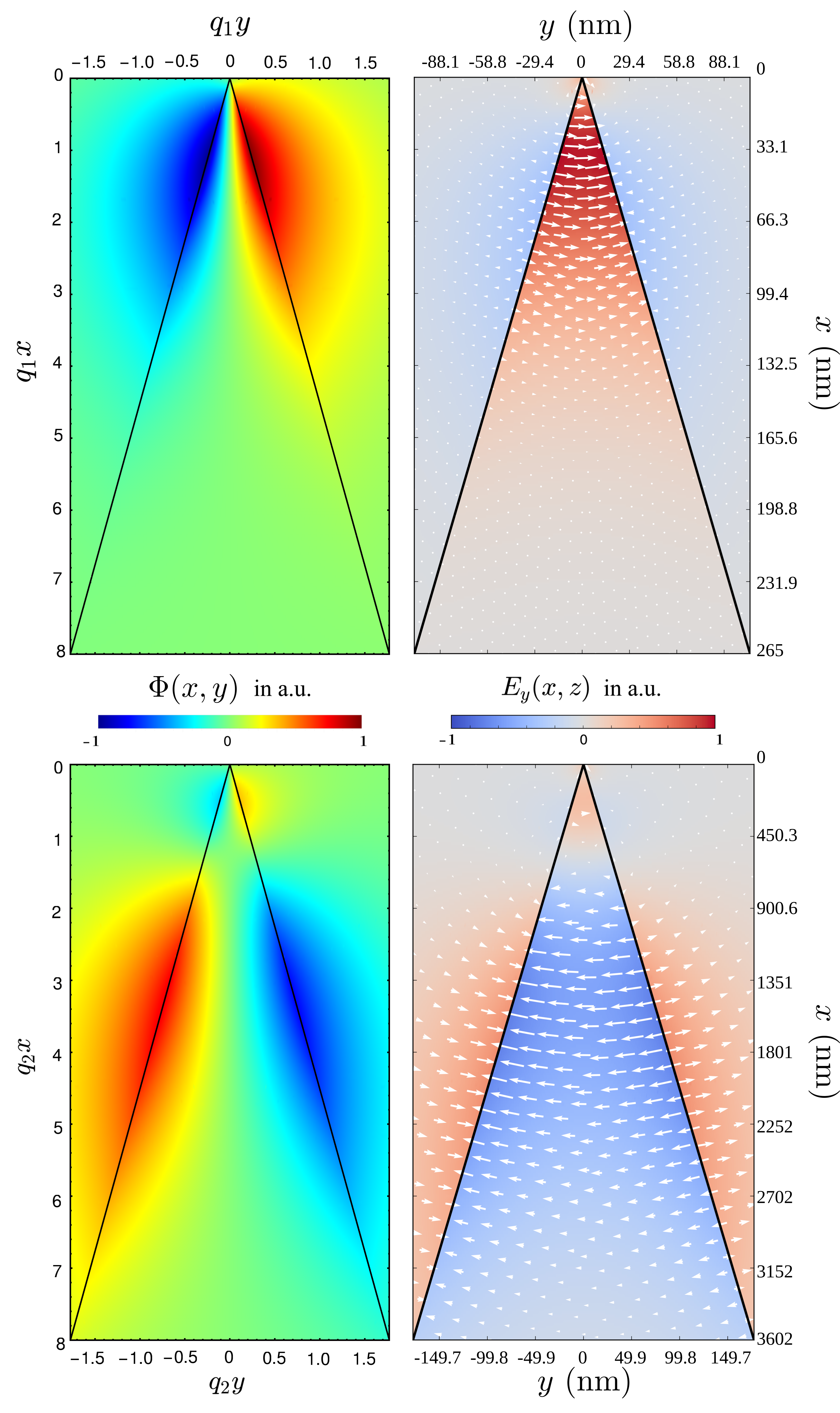}
      \caption{Two-dimensional potential, $\Phi(x,y)$, and electric-field distributions, $\mathbf{E}(x,y) = - \nabla_{2D} \Phi(x,y)$, ascribed to 
      the fundamental (upper panels) and second order (lower panels) WGPs modes, at a resonant frequency of $f=20$ THz. The parameters used 
      in these computations match the ones in Fig.~ \ref{fig:in_plane}. The plotted region is the same in each pair of 2D plots, with axes of 
      the panels on the left written in dimensionless units (normalized to the corresponding propagation constants) and the axes of the plots 
      on the right are given in nanometers.
      }\label{fig:2D_distri}
\end{figure}

We recall that once in possession of the potential evaluated at the graphene, one can build the potential in the entire coordinate space 
using Eq. (\ref{eq:integro_diff_phi_wedge}). From here, the corresponding induced electric field follows directly 
by taking the gradient, i.e. $\mathbf{E}(\mathbf{r}) = - \nabla \Phi(\mathbf{r})$. In what follows, we shall discuss only the 
spatial distributions of the potential and electric field within the plane transverse to the propagation direction, since its 
dependence along the $z$-axis is trivial due to the translational invariance of the system along this direction. The calculated 2D 
distributions (in the $xy$-plane) of the potential and concomitant electric field akin to WGPs 
in a representative $2\vp=25\degree$ dielectric wedge are shown Fig. \ref{fig:2D_distri}.
Note that the electric field in cartesian 
coordinates can be fetched from its polar version by applying a rotation matrix, namely $(E_x,E_y)^T = R(\vp) (E_r,E_\theta)^T$. 
The figure exhibits telling evidence of the remarkable confinement WGP are able to attain near the apex of the wedge (the results for 
the groove are qualitatively similar, albeit with slightly less localization for the same resonant frequency); this can be observed 
both from the induced potential or the electric-field. The intensity plot in the background of the vectorial representation of the 
2D $\mathbf{E}(x,y)$ refers to the value of the $y$-component of this quantity, which is dominant in most of the spatial region. 
Notice that since the potential has odd symmetry with relation to the $y=0$ plane, then the correspondent 
component of the electric field must be of even symmetry. It should 
also be highlighted that the vertical confinement (with respect to the apex edge) decreases rather rapidly as 
we move from the fundamental resonance to the higher order branches of the 
polaritonic spectrum. For the particular case depicted in the figure, 
the lowest energy mode essentially remains localized within the first $200$ nm, whereas for the second order mode that distance 
grows beyond $3.6\,\mu$m. Figure \ref{fig:2D_distri} also serves to support our previous claim 
that despite inherent simplifications in our theory, it 
may still be applied to realistic wedges (grooves) of finite height (depth). More specifically, the results portrayed in the 
figure accurately describe the 1st WGP in a $200$nm-tall dielectric wedge, 
or, instead, \emph{both} WGPs modes in a wedge with about $4\,\mu$m in height.% (with a $25\degree$ opening).

The behavior of the spatial distributions displayed in Fig.~\ref{fig:2D_distri} is qualitatively maintained throughout most 
of the dispersion curve (and similarly for different angles), the only important difference being the degree of confinement 
in the whereabouts of the apex edge. Therefore, an infinitely vast number of V-shaped geometric configurations, with different angles, heights or depths, may be engineered depending on the required level of localization and/or frequency region of interest. In 
this regard, one can use our model to effectively design and architecture a device which meets the pre-established 
requirements in terms of operating frequency range and dimensions.

% ================================================== // ==================================================

% *************************************************************
% ::                      CONCLUSIONS                        ::
% *************************************************************
\section*{Concluding Remarks and Outlook}

In summary, we envision the exploitation of the folding of an otherwise planar graphene sheet---or any other SPP-supporting 
2D material---into an (out-of-plane) triangular-like shape as a mean to achieve deep subwavelength waveguiding and light-localization using the plasmon 
modes guided along the apex of a V-shaped substrate. The fabrication of such devices is well within reach of current experimental 
state-of-the-art capabilities, as much of the technology has already been developed in the context of traditional 3D 
metal plasmonics. As an example, nanoimprint approaches developed for metal grooves\cite{Nielsen:2008} could be 
readily applied also to the present case. The same goes in what concerns potential mechanisms to excite these modes, since the same 
techniques used to excite CPPs could also be applied to excite WGP/GGP modes (e.g. end-fire coupling or excitation by fast electrons).\cite{AsgerRevCPP}
In this article, we have outlined a quasi-analytic method to theoretically 
describe graphene plasmons either guided along the ridge of a dielectric wedge 
or along the valley of a groove carved in an insulating substrate. The model accurately obtains the spectrum of the plasmonic eigenmodes 
supported by the aforementioned structures, as well as the potential and electric field distributions akin to those excitations. 
The computed modal distributions advocate the ability of these modes to achieve large field-enhancements and to deliver strong light-localization 
in the neighborhood of the triangular edge. Interestingly, we have found that the dispersion of 
each particular WGP/GGP mode obeys a universal scaling law, in the sense that 
the functional dependence of the flat-GSP spectrum is maintained up to a multiplication constant that depends solely on the particular 
geometric configuration of the system (i.e., the angle $\vp$ for a given $\ep_1$ and $\ep_2$). The results of our calculations were verified by performing rigorous electrodynamic 
simulations based on the FEM, to which a very good agreement was observed. We note, however, that the quasi-analytical model 
presented above not only provides more physical insight when gauged against numerical simulations, but is also far less computationally 
demanding and less time-consuming that the latter, since all the eigenmodes for a given structure can be determined at once from a single 
computation.
Moreover, we have showed that WGPs and GPPs may be suitable candidates for versatile platforms (specially when 
taking advantage of the gate-tunability of graphene 
plasmons) to effectively route highly confined EM radiation. In this context, subsequent theoretical investigations of wedges/grooves with 
more realistic, rounded edges should constitute the basis of future work. 
The rounding of the apex leads to a slight decrease in the effective index of the plasmon modes, as shown by our finite-element simulations (cf. SI).
Nevertheless, we expect our results to be qualitatively robust with respect to rounding. 
Furthermore, it was demonstrated that even for infinitely sharp metallic apexes, non-local 
effects prevent the emergence of singularities at the tip.\cite{Toscano:13}
The case of a 2D layer deposited on a V-shaped 
substrate with a frequency-dependent dielectric function, e.g. a metal or a polar medium, can also be 
considered using the theory developed here.
In addition, the effects of strain owing to the folding of the 2D sheet 
around the apex remain largely unexplored in what concerns their implications in plasmonic 
excitations [for instance, in electronic transport it is well-known that strain introduces a scattering potential for 
the 2D Dirac fermions (via a gauge field) which can lead to the 
modification of the bandstructure].\cite{Manes:2013,Bruno:2016}
We thus expect that this work will fuel future experimental realizations of WGPs/GGPs, 
as we believe that such modes may hold interesting implications for future all-photonic circuitries at the nanoscale.

% ================================================== // ==================================================
% -------------------------------------------------------------
% #                      Acknowledgments                      #
% -------------------------------------------------------------

\begin{acknowledgement}
\noindent
PADG, SX and NAM acknowledge financial support from the Danish National Research Foundation through the sponsoring of the 
Center for Nanostructured Graphene (CNG), Project DNRF103. NMRP acknowledges funding from the European Commission within of the project 
``Graphene-Driven Revolutions in ICT and Beyond'' (Ref. No. 696656), the Portuguese Foundation for Science and Technology 
(FCT) in the framework of the Strategic Financing UID/FIS/04650/2013,
and the hospitality of the MackGraphe Center, Mackenzie Presbyterian University.
\end{acknowledgement}

\begin{suppinfo}

Supporting Information Available: In the supporting information we describe the derivation of the equations 
given in the main article with utmost detail, while also outlining the calculation of the Green's function akin to a V-shaped interface 
between two insulators. Finally, we give further information about propagation losses, and critically analyze the dependence of the concurrently conducted finite element method (FEM) simulations 
on the radius of the curvature at the triangular apex, since the numerical simulations---contrarily to the analytics---cannot rigorously deal with the 
sharp tip of the V-structure.

\end{suppinfo}

\bibliography{refs_WGPs}

\providecommand{\latin}[1]{#1}
\providecommand*\mcitethebibliography{\thebibliography}
\csname @ifundefined\endcsname{endmcitethebibliography}
  {\let\endmcitethebibliography\endthebibliography}{}
\begin{mcitethebibliography}{84}
\providecommand*\natexlab[1]{#1}
\providecommand*\mciteSetBstSublistMode[1]{}
\providecommand*\mciteSetBstMaxWidthForm[2]{}
\providecommand*\mciteBstWouldAddEndPuncttrue
  {\def\EndOfBibitem{\unskip.}}
\providecommand*\mciteBstWouldAddEndPunctfalse
  {\let\EndOfBibitem\relax}
\providecommand*\mciteSetBstMidEndSepPunct[3]{}
\providecommand*\mciteSetBstSublistLabelBeginEnd[3]{}
\providecommand*\EndOfBibitem{}
\mciteSetBstSublistMode{f}
\mciteSetBstMaxWidthForm{subitem}{(\alph{mcitesubitemcount})}
\mciteSetBstSublistLabelBeginEnd
  {\mcitemaxwidthsubitemform\space}
  {\relax}
  {\relax}

\bibitem[Saleh and Teich(2012)Saleh, and Teich]{Saleh}
Saleh,~B. E.~A.; Teich,~M.~C. \emph{Fundamentals of Photonics}, 2nd ed.; Wiley,
  2012\relax
\mciteBstWouldAddEndPuncttrue
\mciteSetBstMidEndSepPunct{\mcitedefaultmidpunct}
{\mcitedefaultendpunct}{\mcitedefaultseppunct}\relax
\EndOfBibitem
\bibitem[Atwater(2007)]{Atwater}
Atwater,~H.~A. The Promise of Plasmonics. \emph{Sci. Am.} \textbf{2007},
  \emph{296}, 56--62\relax
\mciteBstWouldAddEndPuncttrue
\mciteSetBstMidEndSepPunct{\mcitedefaultmidpunct}
{\mcitedefaultendpunct}{\mcitedefaultseppunct}\relax
\EndOfBibitem
\bibitem[Maradudin \latin{et~al.}(2014)Maradudin, Barnes, and
  Sambles]{MaradudinModern}
Maradudin,~A.~A.; Barnes,~W.~L.; Sambles,~J.~R. \emph{Modern Plasmonics};
  Elsevier, 2014\relax
\mciteBstWouldAddEndPuncttrue
\mciteSetBstMidEndSepPunct{\mcitedefaultmidpunct}
{\mcitedefaultendpunct}{\mcitedefaultseppunct}\relax
\EndOfBibitem
\bibitem[Maier(2007)]{Maier}
Maier,~S.~A. \emph{Plasmonics: Fundamentals and Applications}; Springer: New
  York, 2007\relax
\mciteBstWouldAddEndPuncttrue
\mciteSetBstMidEndSepPunct{\mcitedefaultmidpunct}
{\mcitedefaultendpunct}{\mcitedefaultseppunct}\relax
\EndOfBibitem
\bibitem[Barnes \latin{et~al.}(2003)Barnes, Dereux, and Ebbesen]{Nat424}
Barnes,~W.~L.; Dereux,~A.; Ebbesen,~T.~W. Surface plasmon subwavelength optics.
  \emph{Nature} \textbf{2003}, \emph{424}, 824--830\relax
\mciteBstWouldAddEndPuncttrue
\mciteSetBstMidEndSepPunct{\mcitedefaultmidpunct}
{\mcitedefaultendpunct}{\mcitedefaultseppunct}\relax
\EndOfBibitem
\bibitem[Armstrong(2012)]{difractionfree}
Armstrong,~S. Plasmonics: Diffraction-free surface waves. \emph{Nature Photon.}
  \textbf{2012}, \emph{6}, 720\relax
\mciteBstWouldAddEndPuncttrue
\mciteSetBstMidEndSepPunct{\mcitedefaultmidpunct}
{\mcitedefaultendpunct}{\mcitedefaultseppunct}\relax
\EndOfBibitem
\bibitem[Barnes(2006)]{BarnesJOA}
Barnes,~W.~L. Surface plasmon-polariton length scales: a route to
  sub-wavelength optics. \emph{J. Opt. A: Pure Appl. Opt.} \textbf{2006},
  \emph{8}, S87\relax
\mciteBstWouldAddEndPuncttrue
\mciteSetBstMidEndSepPunct{\mcitedefaultmidpunct}
{\mcitedefaultendpunct}{\mcitedefaultseppunct}\relax
\EndOfBibitem
\bibitem[Gramotnev and Bozhevolnyi(2010)Gramotnev, and
  Bozhevolnyi]{Gramotnev:2010}
Gramotnev,~D.~K.; Bozhevolnyi,~S.~I. Plasmonics beyond the diffraction limit.
  \emph{Nature Photon.} \textbf{2010}, \emph{4}, 83--91\relax
\mciteBstWouldAddEndPuncttrue
\mciteSetBstMidEndSepPunct{\mcitedefaultmidpunct}
{\mcitedefaultendpunct}{\mcitedefaultseppunct}\relax
\EndOfBibitem
\bibitem[Ozbay(2006)]{Sci311}
Ozbay,~E. Plasmonics: Merging Photonics and Electronics at Nanoscale
  Dimensions. \emph{Science} \textbf{2006}, \emph{311}, 189--193\relax
\mciteBstWouldAddEndPuncttrue
\mciteSetBstMidEndSepPunct{\mcitedefaultmidpunct}
{\mcitedefaultendpunct}{\mcitedefaultseppunct}\relax
\EndOfBibitem
\bibitem[Bozhevolnyi \latin{et~al.}(2006)Bozhevolnyi, Volkov, Devaux, Laluet,
  and Ebbesen]{Sergey_nat440}
Bozhevolnyi,~S.~I.; Volkov,~V.~S.; Devaux,~E.; Laluet,~J.-Y.; Ebbesen,~T.~W.
  Channel plasmon subwavelength waveguide components including interferometers
  and ring resonators. \emph{Nature} \textbf{2006}, \emph{440}, 508--511\relax
\mciteBstWouldAddEndPuncttrue
\mciteSetBstMidEndSepPunct{\mcitedefaultmidpunct}
{\mcitedefaultendpunct}{\mcitedefaultseppunct}\relax
\EndOfBibitem
\bibitem[Han and Bozhevolnyi(2013)Han, and Bozhevolnyi]{Sergey_RepProgPhys}
Han,~Z.; Bozhevolnyi,~S.~I. Radiation guiding with surface plasmon polaritons.
  \emph{Rep. Prog. Phys.} \textbf{2013}, \emph{76}, 016402\relax
\mciteBstWouldAddEndPuncttrue
\mciteSetBstMidEndSepPunct{\mcitedefaultmidpunct}
{\mcitedefaultendpunct}{\mcitedefaultseppunct}\relax
\EndOfBibitem
\bibitem[Bozhevolnyi(2008)]{SergeyBook}
Bozhevolnyi,~S.~I. \emph{Plasmonic Nanoguides and Circuits}; Pan Stanford:
  Singapore, 2008\relax
\mciteBstWouldAddEndPuncttrue
\mciteSetBstMidEndSepPunct{\mcitedefaultmidpunct}
{\mcitedefaultendpunct}{\mcitedefaultseppunct}\relax
\EndOfBibitem
\bibitem[Bergman and Stockman(2003)Bergman, and
  Stockman]{PhysRevLett.90.027402}
Bergman,~D.~J.; Stockman,~M.~I. Surface Plasmon Amplification by Stimulated
  Emission of Radiation: Quantum Generation of Coherent Surface Plasmons in
  Nanosystems. \emph{Phys. Rev. Lett.} \textbf{2003}, \emph{90}, 027402\relax
\mciteBstWouldAddEndPuncttrue
\mciteSetBstMidEndSepPunct{\mcitedefaultmidpunct}
{\mcitedefaultendpunct}{\mcitedefaultseppunct}\relax
\EndOfBibitem
\bibitem[Berini and Leon(2012)Berini, and Leon]{nphot6las}
Berini,~P.; Leon,~I.~D. Surface plasmon-polariton amplifiers and lasers.
  \emph{Nature Photon.} \textbf{2012}, \emph{6}, 16--24\relax
\mciteBstWouldAddEndPuncttrue
\mciteSetBstMidEndSepPunct{\mcitedefaultmidpunct}
{\mcitedefaultendpunct}{\mcitedefaultseppunct}\relax
\EndOfBibitem
\bibitem[Ma \latin{et~al.}(2013)Ma, Oulton, Sorger, and Zhang]{LPR7}
Ma,~R.-M.; Oulton,~R.~F.; Sorger,~V.~J.; Zhang,~X. Plasmon lasers: coherent
  light source at molecular scales. \emph{Laser Photon. Rev.} \textbf{2013},
  \emph{7}, 1--21\relax
\mciteBstWouldAddEndPuncttrue
\mciteSetBstMidEndSepPunct{\mcitedefaultmidpunct}
{\mcitedefaultendpunct}{\mcitedefaultseppunct}\relax
\EndOfBibitem
\bibitem[Haffner \latin{et~al.}(2015)Haffner, Heni, Fedoryshyn, Niegemann,
  Melikyan, Elder, Baeuerle, Salamin, Josten, Koch, Hoessbacher, Ducry, Juchli,
  Emboras, Hillerkuss, Kohl, Dalton, Hafner, and Leuthold]{Haffner:2015}
Haffner,~C. \latin{et~al.}  All-plasmonic Mach-Zehnder modulator enabling
  optical high-speed communication at the microscale. \emph{Nature Photon.}
  \textbf{2015}, \emph{9}, 525\relax
\mciteBstWouldAddEndPuncttrue
\mciteSetBstMidEndSepPunct{\mcitedefaultmidpunct}
{\mcitedefaultendpunct}{\mcitedefaultseppunct}\relax
\EndOfBibitem
\bibitem[Chang \latin{et~al.}(2006)Chang, S\o{}rensen, Hemmer, and
  Lukin]{PRL97}
Chang,~D.~E.; S\o{}rensen,~A.~S.; Hemmer,~P.~R.; Lukin,~M.~D. Quantum Optics
  with Surface Plasmons. \emph{Phys. Rev. Lett.} \textbf{2006}, \emph{97},
  053002\relax
\mciteBstWouldAddEndPuncttrue
\mciteSetBstMidEndSepPunct{\mcitedefaultmidpunct}
{\mcitedefaultendpunct}{\mcitedefaultseppunct}\relax
\EndOfBibitem
\bibitem[Heeres \latin{et~al.}(2013)Heeres, Kouwenhoven, and Zwiller]{nnano8}
Heeres,~R.~W.; Kouwenhoven,~L.~P.; Zwiller,~V. Quantum interference in
  plasmonic circuits. \emph{Nature Nanotechnol.} \textbf{2013}, \emph{8},
  719--722\relax
\mciteBstWouldAddEndPuncttrue
\mciteSetBstMidEndSepPunct{\mcitedefaultmidpunct}
{\mcitedefaultendpunct}{\mcitedefaultseppunct}\relax
\EndOfBibitem
\bibitem[Fakonas \latin{et~al.}(2014)Fakonas, Lee, Kelaita, and
  Atwater]{nphoton8}
Fakonas,~J.~S.; Lee,~H.; Kelaita,~Y.~A.; Atwater,~H.~A. Two-plasmon quantum
  interference. \emph{Nature Photon.} \textbf{2014}, \emph{8}, 317--320\relax
\mciteBstWouldAddEndPuncttrue
\mciteSetBstMidEndSepPunct{\mcitedefaultmidpunct}
{\mcitedefaultendpunct}{\mcitedefaultseppunct}\relax
\EndOfBibitem
\bibitem[Tame \latin{et~al.}(2013)Tame, McEnery, Ozdemir, Lee, Maier, and
  Kim]{nphysQP}
Tame,~M.~S.; McEnery,~K.~R.; Ozdemir,~S.~K.; Lee,~J.; Maier,~S.~A.; Kim,~M.~S.
  Quantum plasmonics. \emph{Nature Phys.} \textbf{2013}, \emph{9},
  329--340\relax
\mciteBstWouldAddEndPuncttrue
\mciteSetBstMidEndSepPunct{\mcitedefaultmidpunct}
{\mcitedefaultendpunct}{\mcitedefaultseppunct}\relax
\EndOfBibitem
\bibitem[Bermudez-Urena \latin{et~al.}(2015)Bermudez-Urena,
  Gonzalez-Ballestero, Geiselmann, Marty, Radko, Holmgaard, Alaverdyan, Moreno,
  García-Vidal, Bozhevolnyi, and Quidant]{Bermudez-Urena:2015}
Bermudez-Urena,~E.; Gonzalez-Ballestero,~C.; Geiselmann,~M.; Marty,~R.;
  Radko,~I.~P.; Holmgaard,~T.; Alaverdyan,~Y.; Moreno,~E.;
  García-Vidal,~F.~J.; Bozhevolnyi,~S.~I.; Quidant,~R. Coupling of individual
  quantum emitters to channel plasmons. \emph{Nature Commun.} \textbf{2015},
  \emph{6}, 7883\relax
\mciteBstWouldAddEndPuncttrue
\mciteSetBstMidEndSepPunct{\mcitedefaultmidpunct}
{\mcitedefaultendpunct}{\mcitedefaultseppunct}\relax
\EndOfBibitem
\bibitem[Aćimović \latin{et~al.}(2014)Aćimović, Ortega, Sanz, Berthelot,
  Garcia-Cordero, Renger, Maerkl, Kreuzer, and Quidant]{nl14}
Aćimović,~S.~S.; Ortega,~M.~A.; Sanz,~V.; Berthelot,~J.;
  Garcia-Cordero,~J.~L.; Renger,~J.; Maerkl,~S.~J.; Kreuzer,~M.~P.; Quidant,~R.
  LSPR Chip for Parallel, Rapid, and Sensitive Detection of Cancer Markers in
  Serum. \emph{Nano Lett.} \textbf{2014}, \emph{14}, 2636--2641\relax
\mciteBstWouldAddEndPuncttrue
\mciteSetBstMidEndSepPunct{\mcitedefaultmidpunct}
{\mcitedefaultendpunct}{\mcitedefaultseppunct}\relax
\EndOfBibitem
\bibitem[Haes \latin{et~al.}(2005)Haes, Haynes, McFarland, Schatz, Duyne, and
  Zou]{mrs30}
Haes,~A.~J.; Haynes,~C.~L.; McFarland,~A.~D.; Schatz,~G.~C.; Duyne,~R. P.~V.;
  Zou,~S. Plasmonic Materials for Surface-Enhanced Sensing and Spectroscopy.
  \emph{MRS Bulletin} \textbf{2005}, \emph{30}, 368--375\relax
\mciteBstWouldAddEndPuncttrue
\mciteSetBstMidEndSepPunct{\mcitedefaultmidpunct}
{\mcitedefaultendpunct}{\mcitedefaultseppunct}\relax
\EndOfBibitem
\bibitem[Brolo(2012)]{nphot6}
Brolo,~A.~G. Plasmonics for future biosensors. \emph{Nature Photon.}
  \textbf{2012}, \emph{6}, 709--713\relax
\mciteBstWouldAddEndPuncttrue
\mciteSetBstMidEndSepPunct{\mcitedefaultmidpunct}
{\mcitedefaultendpunct}{\mcitedefaultseppunct}\relax
\EndOfBibitem
\bibitem[Long and Jing(2014)Long, and Jing]{bioBook}
Long,~Y.-T.; Jing,~C. \emph{Localized Surface Plasmon Resonance Based
  Nanobiosensors}; Springer, 2014\relax
\mciteBstWouldAddEndPuncttrue
\mciteSetBstMidEndSepPunct{\mcitedefaultmidpunct}
{\mcitedefaultendpunct}{\mcitedefaultseppunct}\relax
\EndOfBibitem
\bibitem[Rodrigo \latin{et~al.}(2015)Rodrigo, Limaj, DavideJanner, Etezadi,
  García~de Abajo, Pruneri, and Altug]{Rodrigo10072015}
Rodrigo,~D.; Limaj,~O.; DavideJanner,; Etezadi,~D.; García~de Abajo,~F.~J.;
  Pruneri,~V.; Altug,~H. Mid-infrared plasmonic biosensing with graphene.
  \emph{Science} \textbf{2015}, \emph{349}, 165--168\relax
\mciteBstWouldAddEndPuncttrue
\mciteSetBstMidEndSepPunct{\mcitedefaultmidpunct}
{\mcitedefaultendpunct}{\mcitedefaultseppunct}\relax
\EndOfBibitem
\bibitem[Fang \latin{et~al.}(2005)Fang, Lee, Sun, and Zhang]{Fang534}
Fang,~N.; Lee,~H.; Sun,~C.; Zhang,~X. Sub-Diffraction-Limited Optical Imaging
  with a Silver Superlens. \emph{Science} \textbf{2005}, \emph{308},
  534--537\relax
\mciteBstWouldAddEndPuncttrue
\mciteSetBstMidEndSepPunct{\mcitedefaultmidpunct}
{\mcitedefaultendpunct}{\mcitedefaultseppunct}\relax
\EndOfBibitem
\bibitem[Liu \latin{et~al.}(2011)Liu, Hentschel, Weiss, Alivisatos, and
  Giessen]{Liu1407}
Liu,~N.; Hentschel,~M.; Weiss,~T.; Alivisatos,~A.~P.; Giessen,~H.
  Three-Dimensional Plasmon Rulers. \emph{Science} \textbf{2011}, \emph{332},
  1407--1410\relax
\mciteBstWouldAddEndPuncttrue
\mciteSetBstMidEndSepPunct{\mcitedefaultmidpunct}
{\mcitedefaultendpunct}{\mcitedefaultseppunct}\relax
\EndOfBibitem
\bibitem[Vo-Dinh \latin{et~al.}(2013)Vo-Dinh, Fales, Griffin, Khoury, Liu, Ngo,
  Norton, Register, Wang, and Yuan]{med}
Vo-Dinh,~T.; Fales,~A.~M.; Griffin,~G.~D.; Khoury,~C.~G.; Liu,~Y.; Ngo,~H.;
  Norton,~S.~J.; Register,~J.~K.; Wang,~H.-N.; Yuan,~H. Plasmonic nanoprobes:
  from chemical sensing to medical diagnostics and therapy. \emph{Nanoscale}
  \textbf{2013}, \emph{5}, 10127--10140\relax
\mciteBstWouldAddEndPuncttrue
\mciteSetBstMidEndSepPunct{\mcitedefaultmidpunct}
{\mcitedefaultendpunct}{\mcitedefaultseppunct}\relax
\EndOfBibitem
\bibitem[Zheng \latin{et~al.}(2012)Zheng, Kiraly, Weiss, and Huang]{med2}
Zheng,~Y.~B.; Kiraly,~B.; Weiss,~P.~S.; Huang,~T.~J. Molecular plasmonics for
  biology and nanomedicine. \emph{Nanomedicine} \textbf{2012}, \emph{7}, 751 --
  770\relax
\mciteBstWouldAddEndPuncttrue
\mciteSetBstMidEndSepPunct{\mcitedefaultmidpunct}
{\mcitedefaultendpunct}{\mcitedefaultseppunct}\relax
\EndOfBibitem
\bibitem[Bardhan \latin{et~al.}(2011)Bardhan, Lal, Joshi, and
  Halas]{HalasThera}
Bardhan,~R.; Lal,~S.; Joshi,~A.; Halas,~N.~J. Theranostic Nanoshells: From
  Probe Design to Imaging and Treatment of Cancer. \emph{Accounts of Chemical
  Research} \textbf{2011}, \emph{44}, 936--946\relax
\mciteBstWouldAddEndPuncttrue
\mciteSetBstMidEndSepPunct{\mcitedefaultmidpunct}
{\mcitedefaultendpunct}{\mcitedefaultseppunct}\relax
\EndOfBibitem
\bibitem[Fang and Sun(2015)Fang, and Sun]{PlasmWG}
Fang,~Y.; Sun,~M. Nanoplasmonic waveguides: towards applications in integrated
  nanophotonic circuits. \emph{Light Sci. Appl.} \textbf{2015}, \emph{4},
  e294\relax
\mciteBstWouldAddEndPuncttrue
\mciteSetBstMidEndSepPunct{\mcitedefaultmidpunct}
{\mcitedefaultendpunct}{\mcitedefaultseppunct}\relax
\EndOfBibitem
\bibitem[Krenn(2003)]{NPsWVG}
Krenn,~J.~R. Nanoparticle waveguides: Watching energy transfer. \emph{Nature
  Mater.} \textbf{2003}, \emph{2}, 210--211\relax
\mciteBstWouldAddEndPuncttrue
\mciteSetBstMidEndSepPunct{\mcitedefaultmidpunct}
{\mcitedefaultendpunct}{\mcitedefaultseppunct}\relax
\EndOfBibitem
\bibitem[Maier \latin{et~al.}(2003)Maier, Kik, Atwater, Meltzer, Harel, Koel,
  and Requicha]{NPsWVG2}
Maier,~S.~A.; Kik,~P.~G.; Atwater,~H.~A.; Meltzer,~S.; Harel,~E.; Koel,~B.~E.;
  Requicha,~A. A.~G. Local detection of electromagnetic energy transport below
  the diffraction limit in metal nanoparticle plasmon waveguides. \emph{Nature
  Mater.} \textbf{2003}, \emph{2}, 229--232\relax
\mciteBstWouldAddEndPuncttrue
\mciteSetBstMidEndSepPunct{\mcitedefaultmidpunct}
{\mcitedefaultendpunct}{\mcitedefaultseppunct}\relax
\EndOfBibitem
\bibitem[Alù \latin{et~al.}(2011)Alù, Belov, and Engheta]{guidingNPs}
Alù,~A.; Belov,~P.~A.; Engheta,~N. Coupling and guided propagation along
  parallel chains of plasmonic nanoparticles. \emph{New J. Phys.}
  \textbf{2011}, \emph{13}, 033026\relax
\mciteBstWouldAddEndPuncttrue
\mciteSetBstMidEndSepPunct{\mcitedefaultmidpunct}
{\mcitedefaultendpunct}{\mcitedefaultseppunct}\relax
\EndOfBibitem
\bibitem[Smith \latin{et~al.}(2015)Smith, Stenger, Kristensen, Mortensen, and
  Bozhevolnyi]{AsgerRevCPP}
Smith,~C. L.~C.; Stenger,~N.; Kristensen,~A.; Mortensen,~N.~A.;
  Bozhevolnyi,~S.~I. Gap and channeled plasmons in tapered grooves: a review.
  \emph{Nanoscale} \textbf{2015}, \emph{7}, 9355--9386\relax
\mciteBstWouldAddEndPuncttrue
\mciteSetBstMidEndSepPunct{\mcitedefaultmidpunct}
{\mcitedefaultendpunct}{\mcitedefaultseppunct}\relax
\EndOfBibitem
\bibitem[Gramotnev and Bozhevolnyi(2014)Gramotnev, and Bozhevolnyi]{nanoFocus}
Gramotnev,~D.~K.; Bozhevolnyi,~S.~I. Nanofocusing of electromagnetic radiation.
  \emph{Nature Photon.} \textbf{2014}, \emph{9}, 13--22\relax
\mciteBstWouldAddEndPuncttrue
\mciteSetBstMidEndSepPunct{\mcitedefaultmidpunct}
{\mcitedefaultendpunct}{\mcitedefaultseppunct}\relax
\EndOfBibitem
\bibitem[Raza \latin{et~al.}(2014)Raza, Stenger, Pors, Holmgaard,
  Kadkhodazadeh, Wagner, Pedersen, Wubs, Bozhevolnyi, and Mortensen]{raza:2014}
Raza,~S.; Stenger,~N.; Pors,~A.; Holmgaard,~T.; Kadkhodazadeh,~S.;
  Wagner,~J.~B.; Pedersen,~K.; Wubs,~M.; Bozhevolnyi,~S.~I.; Mortensen,~N.~A.
  Extremely confined gap surface-plasmon modes excited by electrons.
  \emph{Nature Commun.} \textbf{2014}, \emph{5}, 4125\relax
\mciteBstWouldAddEndPuncttrue
\mciteSetBstMidEndSepPunct{\mcitedefaultmidpunct}
{\mcitedefaultendpunct}{\mcitedefaultseppunct}\relax
\EndOfBibitem
\bibitem[Pile and Gramotnev(2004)Pile, and Gramotnev]{Pile:04}
Pile,~D. F.~P.; Gramotnev,~D.~K. Channel plasmon--polariton in a triangular
  groove on a metal surface. \emph{Opt. Lett.} \textbf{2004}, \emph{29},
  1069--1071\relax
\mciteBstWouldAddEndPuncttrue
\mciteSetBstMidEndSepPunct{\mcitedefaultmidpunct}
{\mcitedefaultendpunct}{\mcitedefaultseppunct}\relax
\EndOfBibitem
\bibitem[Moreno \latin{et~al.}(2006)Moreno, García-Vidal, Rodrigo,
  Martin-Moreno, and Bozhevolnyi]{Moreno:06}
Moreno,~E.; García-Vidal,~F.~J.; Rodrigo,~S.~G.; Martin-Moreno,~L.;
  Bozhevolnyi,~S.~I. Channel plasmon-polaritons: modal shape, dispersion, and
  losses. \emph{Opt. Lett.} \textbf{2006}, \emph{31}, 3447--3449\relax
\mciteBstWouldAddEndPuncttrue
\mciteSetBstMidEndSepPunct{\mcitedefaultmidpunct}
{\mcitedefaultendpunct}{\mcitedefaultseppunct}\relax
\EndOfBibitem
\bibitem[Li \latin{et~al.}(2013)Li, Jiang, Shen, and Deng]{telecom}
Li,~X.; Jiang,~T.; Shen,~L.; Deng,~X. Subwavelength guiding of channel plasmon
  polaritons by textured metallic grooves at telecom wavelengths. \emph{Appl.
  Phys. Lett.} \textbf{2013}, \emph{102}, 031606\relax
\mciteBstWouldAddEndPuncttrue
\mciteSetBstMidEndSepPunct{\mcitedefaultmidpunct}
{\mcitedefaultendpunct}{\mcitedefaultseppunct}\relax
\EndOfBibitem
\bibitem[Volkov \latin{et~al.}(2006)Volkov, Bozhevolnyi, Devaux, and
  Ebbesen]{Volkov:06}
Volkov,~V.~S.; Bozhevolnyi,~S.~I.; Devaux,~E.; Ebbesen,~T.~W. Compact gradual
  bends for channel plasmon polaritons. \emph{Opt. Express} \textbf{2006},
  \emph{14}, 4494--4503\relax
\mciteBstWouldAddEndPuncttrue
\mciteSetBstMidEndSepPunct{\mcitedefaultmidpunct}
{\mcitedefaultendpunct}{\mcitedefaultseppunct}\relax
\EndOfBibitem
\bibitem[Dobrzynski and Maradudin(1972)Dobrzynski, and
  Maradudin]{PhysRevB.6.3810}
Dobrzynski,~L.; Maradudin,~A.~A. Electrostatic Edge Modes in a Dielectric
  Wedge. \emph{Phys. Rev. B} \textbf{1972}, \emph{6}, 3810--3815\relax
\mciteBstWouldAddEndPuncttrue
\mciteSetBstMidEndSepPunct{\mcitedefaultmidpunct}
{\mcitedefaultendpunct}{\mcitedefaultseppunct}\relax
\EndOfBibitem
\bibitem[Boardman \latin{et~al.}(1981)Boardman, Aers, and
  Teshima]{PhysRevB.24.5703}
Boardman,~A.~D.; Aers,~G.~C.; Teshima,~R. Retarded edge modes of a parabolic
  wedge. \emph{Phys. Rev. B} \textbf{1981}, \emph{24}, 5703--5712\relax
\mciteBstWouldAddEndPuncttrue
\mciteSetBstMidEndSepPunct{\mcitedefaultmidpunct}
{\mcitedefaultendpunct}{\mcitedefaultseppunct}\relax
\EndOfBibitem
\bibitem[Boardman \latin{et~al.}(1985)Boardman, Garcia-Molina, Gras-Marti, and
  Louis]{PhysRevB.32.6045}
Boardman,~A.~D.; Garcia-Molina,~R.; Gras-Marti,~A.; Louis,~E. Electrostatic
  edge modes of a hyperbolic dielectric wedge: Analytical solution. \emph{Phys.
  Rev. B} \textbf{1985}, \emph{32}, 6045--6047\relax
\mciteBstWouldAddEndPuncttrue
\mciteSetBstMidEndSepPunct{\mcitedefaultmidpunct}
{\mcitedefaultendpunct}{\mcitedefaultseppunct}\relax
\EndOfBibitem
\bibitem[Novikov and Maradudin(2002)Novikov, and Maradudin]{PhysRevB.66.035403}
Novikov,~I.~V.; Maradudin,~A.~A. Channel polaritons. \emph{Phys. Rev. B}
  \textbf{2002}, \emph{66}, 035403\relax
\mciteBstWouldAddEndPuncttrue
\mciteSetBstMidEndSepPunct{\mcitedefaultmidpunct}
{\mcitedefaultendpunct}{\mcitedefaultseppunct}\relax
\EndOfBibitem
\bibitem[Lu and Maradudin(1990)Lu, and Maradudin]{PhysRevB.42.11159}
Lu,~J.~Q.; Maradudin,~A.~A. Channel plasmons. \emph{Phys. Rev. B}
  \textbf{1990}, \emph{42}, 11159--11165\relax
\mciteBstWouldAddEndPuncttrue
\mciteSetBstMidEndSepPunct{\mcitedefaultmidpunct}
{\mcitedefaultendpunct}{\mcitedefaultseppunct}\relax
\EndOfBibitem
\bibitem[Volkov \latin{et~al.}(2009)Volkov, Bozhevolnyi, Rodrigo,
  Martín-Moreno, García-Vidal, Devaux, and Ebbesen]{nl9_nanoFocus}
Volkov,~V.~S.; Bozhevolnyi,~S.~I.; Rodrigo,~S.~G.; Martín-Moreno,~L.;
  García-Vidal,~F.~J.; Devaux,~E.; Ebbesen,~T.~W. Nanofocusing with Channel
  Plasmon Polaritons. \emph{Nano Lett.} \textbf{2009}, \emph{9},
  1278--1282\relax
\mciteBstWouldAddEndPuncttrue
\mciteSetBstMidEndSepPunct{\mcitedefaultmidpunct}
{\mcitedefaultendpunct}{\mcitedefaultseppunct}\relax
\EndOfBibitem
\bibitem[Bozhevolnyi and Nerkararyan(2009)Bozhevolnyi, and
  Nerkararyan]{Bozhevolnyi_oe17}
Bozhevolnyi,~S.~I.; Nerkararyan,~K.~V. Channel plasmon polaritons guided by
  graded gaps: closed-form solutions. \emph{Opt. Express} \textbf{2009},
  \emph{17}, 10327--10334\relax
\mciteBstWouldAddEndPuncttrue
\mciteSetBstMidEndSepPunct{\mcitedefaultmidpunct}
{\mcitedefaultendpunct}{\mcitedefaultseppunct}\relax
\EndOfBibitem
\bibitem[Bozhevolnyi and Nerkararyan(2009)Bozhevolnyi, and
  Nerkararyan]{Bozhevolnyi:09}
Bozhevolnyi,~S.~I.; Nerkararyan,~K.~V. Analytic description of channel plasmon
  polaritons. \emph{Opt. Lett.} \textbf{2009}, \emph{34}, 2039--2041\relax
\mciteBstWouldAddEndPuncttrue
\mciteSetBstMidEndSepPunct{\mcitedefaultmidpunct}
{\mcitedefaultendpunct}{\mcitedefaultseppunct}\relax
\EndOfBibitem
\bibitem[Luo \latin{et~al.}(2010)Luo, Pendry, and Aubry]{nl10:Luo}
Luo,~Y.; Pendry,~J.~B.; Aubry,~A. Surface Plasmons and Singularities.
  \emph{Nano Letters} \textbf{2010}, \emph{10}, 4186--4191\relax
\mciteBstWouldAddEndPuncttrue
\mciteSetBstMidEndSepPunct{\mcitedefaultmidpunct}
{\mcitedefaultendpunct}{\mcitedefaultseppunct}\relax
\EndOfBibitem
\bibitem[Geim(2009)]{Geim09}
Geim,~A.~K. Graphene: Status and Prospects. \emph{Science} \textbf{2009},
  \emph{324}, 1530--1534\relax
\mciteBstWouldAddEndPuncttrue
\mciteSetBstMidEndSepPunct{\mcitedefaultmidpunct}
{\mcitedefaultendpunct}{\mcitedefaultseppunct}\relax
\EndOfBibitem
\bibitem[{Castro Neto} \latin{et~al.}(2009){Castro Neto}, Guinea, Peres,
  Novoselov, and Geim]{RMP81}
{Castro Neto},~A.~H.; Guinea,~F.; Peres,~N. M.~R.; Novoselov,~K.~S.;
  Geim,~A.~K. The electronic properties of graphene. \emph{Rev. Mod. Phys.}
  \textbf{2009}, \emph{81}\relax
\mciteBstWouldAddEndPuncttrue
\mciteSetBstMidEndSepPunct{\mcitedefaultmidpunct}
{\mcitedefaultendpunct}{\mcitedefaultseppunct}\relax
\EndOfBibitem
\bibitem[Gon\c{c}alves and Peres(2016)Gon\c{c}alves, and Peres]{GoncalvesPeres}
Gon\c{c}alves,~P. A.~D.; Peres,~N. M.~R. \emph{An Introduction to Graphene
  Plasmonics}; World Scientific: Singapore, 2016\relax
\mciteBstWouldAddEndPuncttrue
\mciteSetBstMidEndSepPunct{\mcitedefaultmidpunct}
{\mcitedefaultendpunct}{\mcitedefaultseppunct}\relax
\EndOfBibitem
\bibitem[García~de Abajo(2014)]{AbajoACSP}
García~de Abajo,~F.~J. Graphene Plasmonics: Challenges and Opportunities.
  \emph{ACS Photonics} \textbf{2014}, \emph{1}, 135--152\relax
\mciteBstWouldAddEndPuncttrue
\mciteSetBstMidEndSepPunct{\mcitedefaultmidpunct}
{\mcitedefaultendpunct}{\mcitedefaultseppunct}\relax
\EndOfBibitem
\bibitem[Xiao \latin{et~al.}(2016)Xiao, Zhu, Li, and Mortensen]{Xiao2016}
Xiao,~S.; Zhu,~X.; Li,~B.-H.; Mortensen,~N.~A. Graphene-plasmon polaritons:
  From fundamental properties to potential applications. \emph{Front. Phys.}
  \textbf{2016}, \emph{11}, 117801\relax
\mciteBstWouldAddEndPuncttrue
\mciteSetBstMidEndSepPunct{\mcitedefaultmidpunct}
{\mcitedefaultendpunct}{\mcitedefaultseppunct}\relax
\EndOfBibitem
\bibitem[Bludov \latin{et~al.}(2013)Bludov, Ferreira, Peres, and
  Vasilevskiy]{Primer}
Bludov,~Y.~V.; Ferreira,~A.; Peres,~N. M.~R.; Vasilevskiy,~M.~I. A Primer on
  Surface Plasmon-Polaritons in Graphene. \emph{Int. J. Mod. Phys. B}
  \textbf{2013}, \emph{27}, 1341001\relax
\mciteBstWouldAddEndPuncttrue
\mciteSetBstMidEndSepPunct{\mcitedefaultmidpunct}
{\mcitedefaultendpunct}{\mcitedefaultseppunct}\relax
\EndOfBibitem
\bibitem[Low and Avouris(2014)Low, and Avouris]{ACSgp}
Low,~T.; Avouris,~P. Graphene Plasmonics for Terahertz to Mid-Infrared
  Applications. \emph{ACS Nano} \textbf{2014}, \emph{8}, 1086--1101\relax
\mciteBstWouldAddEndPuncttrue
\mciteSetBstMidEndSepPunct{\mcitedefaultmidpunct}
{\mcitedefaultendpunct}{\mcitedefaultseppunct}\relax
\EndOfBibitem
\bibitem[Koppens \latin{et~al.}(2011)Koppens, Chang, and García~de
  Abajo]{nlgp}
Koppens,~F. H.~L.; Chang,~D.~E.; García~de Abajo,~F.~J. Graphene Plasmonics: A
  Platform for Strong Light-Matter Interactions. \emph{Nano Lett.}
  \textbf{2011}, \emph{11}, 3370--3377\relax
\mciteBstWouldAddEndPuncttrue
\mciteSetBstMidEndSepPunct{\mcitedefaultmidpunct}
{\mcitedefaultendpunct}{\mcitedefaultseppunct}\relax
\EndOfBibitem
\bibitem[Ju \latin{et~al.}(2011)Ju, Geng, Horng, Girit, Martin, Hao, Bechtel,
  Liang, Zettl, Shen, and Wang]{NatNano}
Ju,~L.; Geng,~B.; Horng,~J.; Girit,~C.; Martin,~M.; Hao,~Z.; Bechtel,~H.~A.;
  Liang,~X.; Zettl,~A.; Shen,~Y.~R.; Wang,~F. Graphene plasmonics for tunable
  terahertz metamaterials. \emph{Nature Nanotechnol.} \textbf{2011}, \emph{6},
  630--634\relax
\mciteBstWouldAddEndPuncttrue
\mciteSetBstMidEndSepPunct{\mcitedefaultmidpunct}
{\mcitedefaultendpunct}{\mcitedefaultseppunct}\relax
\EndOfBibitem
\bibitem[Yan \latin{et~al.}(2013)Yan, Low, Zhu, Wu, Freitag, Li, Guinea,
  Avouris, and Xia]{nphoton7}
Yan,~H.; Low,~T.; Zhu,~W.; Wu,~Y.; Freitag,~M.; Li,~X.; Guinea,~F.;
  Avouris,~P.; Xia,~F. Damping pathways of mid-infrared plasmons in graphene
  nanostructures. \emph{Nature Photon.} \textbf{2013}, \emph{7}, 394--399\relax
\mciteBstWouldAddEndPuncttrue
\mciteSetBstMidEndSepPunct{\mcitedefaultmidpunct}
{\mcitedefaultendpunct}{\mcitedefaultseppunct}\relax
\EndOfBibitem
\bibitem[Luxmoore \latin{et~al.}(2014)Luxmoore, Gan, Liu, Valmorra, Li, Faist,
  and Nash]{luxmoore14}
Luxmoore,~I.~J.; Gan,~C.~H.; Liu,~P.~Q.; Valmorra,~F.; Li,~P.; Faist,~J.;
  Nash,~G.~R. Strong coupling in the far-infrared between graphene plasmons and
  the surface optical phonons of silicon dioxide. \emph{ACS Photonics}
  \textbf{2014}, \emph{1}, 1151\relax
\mciteBstWouldAddEndPuncttrue
\mciteSetBstMidEndSepPunct{\mcitedefaultmidpunct}
{\mcitedefaultendpunct}{\mcitedefaultseppunct}\relax
\EndOfBibitem
\bibitem[Hu \latin{et~al.}(2016)Hu, Yang, Zhai, Hu, Liu, Liu, Sun, and
  Dai]{natcomm_Gribbons}
Hu,~H.; Yang,~X.; Zhai,~F.; Hu,~D.; Liu,~R.; Liu,~K.; Sun,~Z.; Dai,~Q.
  Far-field nanoscale infrared spectroscopy of vibrational fingerprints of
  molecules with graphene plasmons. \emph{Nature Commun.} \textbf{2016},
  \emph{7}, 1151\relax
\mciteBstWouldAddEndPuncttrue
\mciteSetBstMidEndSepPunct{\mcitedefaultmidpunct}
{\mcitedefaultendpunct}{\mcitedefaultseppunct}\relax
\EndOfBibitem
\bibitem[Yan \latin{et~al.}(2012)Yan, Xia, Li, and Avouris]{NJP14}
Yan,~H.; Xia,~F.; Li,~Z.; Avouris,~P. Plasmonics of coupled graphene
  micro-structures. \emph{New J. Phys.} \textbf{2012}, \emph{14}, 125001\relax
\mciteBstWouldAddEndPuncttrue
\mciteSetBstMidEndSepPunct{\mcitedefaultmidpunct}
{\mcitedefaultendpunct}{\mcitedefaultseppunct}\relax
\EndOfBibitem
\bibitem[Fang \latin{et~al.}(2013)Fang, Thongrattanasiri, Schlather, Liu, Ma,
  Wang, Ajayan, Nordlander, Halas, and García~de Abajo]{ACS7}
Fang,~Z.; Thongrattanasiri,~S.; Schlather,~A.; Liu,~Z.; Ma,~L.; Wang,~Y.;
  Ajayan,~P.~M.; Nordlander,~P.; Halas,~N.~J.; García~de Abajo,~F.~J. Gated
  Tunability and Hybridization of Localized Plasmons in Nanostructured
  Graphene. \emph{ACS Nano} \textbf{2013}, \emph{7}, 2388--2395\relax
\mciteBstWouldAddEndPuncttrue
\mciteSetBstMidEndSepPunct{\mcitedefaultmidpunct}
{\mcitedefaultendpunct}{\mcitedefaultseppunct}\relax
\EndOfBibitem
\bibitem[Fang \latin{et~al.}(2014)Fang, Wang, Schlather, Liu, Ajayan,
  García~de Abajo, Nordlander, Zhu, and Halas]{ZFnl14}
Fang,~Z.; Wang,~Y.; Schlather,~A.~E.; Liu,~Z.; Ajayan,~P.~M.; García~de
  Abajo,~F.~J.; Nordlander,~P.; Zhu,~X.; Halas,~N.~J. Active Tunable Absorption
  Enhancement with Graphene Nanodisk Arrays. \emph{Nano Lett.} \textbf{2014},
  \emph{14}, 299--304\relax
\mciteBstWouldAddEndPuncttrue
\mciteSetBstMidEndSepPunct{\mcitedefaultmidpunct}
{\mcitedefaultendpunct}{\mcitedefaultseppunct}\relax
\EndOfBibitem
\bibitem[Zhu \latin{et~al.}(2014)Zhu, Wang, Yan, Larsen, Bøggild, Pedersen,
  Xiao, Zi, and Mortensen]{nl14XZ}
Zhu,~X.; Wang,~W.; Yan,~W.; Larsen,~M.~B.; Bøggild,~P.; Pedersen,~T.~G.;
  Xiao,~S.; Zi,~J.; Mortensen,~N.~A. Plasmon-Phonon Coupling in Large-Area
  Graphene Dot and Antidot Arrays Fabricated by Nanosphere Lithography.
  \emph{Nano Lett.} \textbf{2014}, \emph{14}, 2907--2913\relax
\mciteBstWouldAddEndPuncttrue
\mciteSetBstMidEndSepPunct{\mcitedefaultmidpunct}
{\mcitedefaultendpunct}{\mcitedefaultseppunct}\relax
\EndOfBibitem
\bibitem[Wang \latin{et~al.}(2016)Wang, Li, Almdal, Mortensen, Xiao, and
  Ndoni]{AsgerTinyDisks}
Wang,~Z.; Li,~T.; Almdal,~K.; Mortensen,~N.~A.; Xiao,~S.; Ndoni,~S. Pushing
  graphene plasmon polaritons to the near-infrared window by block copolymer
  nanolithography. \emph{arXiv:1606.02471 [physics.optics]} \textbf{2016},
  \relax
\mciteBstWouldAddEndPunctfalse
\mciteSetBstMidEndSepPunct{\mcitedefaultmidpunct}
{}{\mcitedefaultseppunct}\relax
\EndOfBibitem
\bibitem[Liu \latin{et~al.}(2015)Liu, Valmorra, Maissen, and Faist]{Optica2}
Liu,~P.~Q.; Valmorra,~F.; Maissen,~C.; Faist,~J. Electrically tunable graphene
  anti-dot array terahertz plasmonic crystals exhibiting multi-band resonances.
  \emph{Optica} \textbf{2015}, \emph{2}, 135--140\relax
\mciteBstWouldAddEndPuncttrue
\mciteSetBstMidEndSepPunct{\mcitedefaultmidpunct}
{\mcitedefaultendpunct}{\mcitedefaultseppunct}\relax
\EndOfBibitem
\bibitem[Yeung \latin{et~al.}(2014)Yeung, Chee, Yoon, Song, Kong, and
  Ham]{PlasmonicCrystals}
Yeung,~K. Y.~M.; Chee,~J.; Yoon,~H.; Song,~Y.; Kong,~J.; Ham,~D. Far-Infrared
  Graphene Plasmonic Crystals for Plasmonic Band Engineering. \emph{Nano Lett.}
  \textbf{2014}, \emph{14}, 2479--2484\relax
\mciteBstWouldAddEndPuncttrue
\mciteSetBstMidEndSepPunct{\mcitedefaultmidpunct}
{\mcitedefaultendpunct}{\mcitedefaultseppunct}\relax
\EndOfBibitem
\bibitem[Liu \latin{et~al.}(2013)Liu, Zhang, Ma, Cai, Wang, and Xu]{Liu:13}
Liu,~P.; Zhang,~X.; Ma,~Z.; Cai,~W.; Wang,~L.; Xu,~J. Surface plasmon modes in
  graphene wedge and groove waveguides. \emph{Opt. Express} \textbf{2013},
  \emph{21}, 32432--32440\relax
\mciteBstWouldAddEndPuncttrue
\mciteSetBstMidEndSepPunct{\mcitedefaultmidpunct}
{\mcitedefaultendpunct}{\mcitedefaultseppunct}\relax
\EndOfBibitem
\bibitem[Kim \latin{et~al.}(2011)Kim, Lee, Malone, Chan, Alem\'an, Regan,
  Gannett, Crommie, Cohen, and Zettl]{PhysRevB.83.245433}
Kim,~K.; Lee,~Z.; Malone,~B.~D.; Chan,~K.~T.; Alem\'an,~B.; Regan,~W.;
  Gannett,~W.; Crommie,~M.~F.; Cohen,~M.~L.; Zettl,~A. Multiply folded
  graphene. \emph{Phys. Rev. B} \textbf{2011}, \emph{83}, 245433\relax
\mciteBstWouldAddEndPuncttrue
\mciteSetBstMidEndSepPunct{\mcitedefaultmidpunct}
{\mcitedefaultendpunct}{\mcitedefaultseppunct}\relax
\EndOfBibitem
\bibitem[Pakhnevich \latin{et~al.}(2015)Pakhnevich, Golod, and
  Prinz]{V-graphene}
Pakhnevich,~A.~A.; Golod,~S.~V.; Prinz,~V.~Y. Surface melting of copper during
  graphene growth by chemical vapour deposition. \emph{J. Phys. D: Appl. Phys.}
  \textbf{2015}, \emph{48}, 435303\relax
\mciteBstWouldAddEndPuncttrue
\mciteSetBstMidEndSepPunct{\mcitedefaultmidpunct}
{\mcitedefaultendpunct}{\mcitedefaultseppunct}\relax
\EndOfBibitem
\bibitem[Song \latin{et~al.}(2016)Song, An, Chen, Peng, Zang, and
  Yang]{C6NR01992G}
Song,~Q.; An,~M.; Chen,~X.; Peng,~Z.; Zang,~J.; Yang,~N. Adjustable thermal
  resistor by reversibly folding a graphene sheet. \emph{Nanoscale}
  \textbf{2016}, \relax
\mciteBstWouldAddEndPunctfalse
\mciteSetBstMidEndSepPunct{\mcitedefaultmidpunct}
{}{\mcitedefaultseppunct}\relax
\EndOfBibitem
\bibitem[Zhu \latin{et~al.}(2012)Zhu, Low, Perebeinos, Bol, Zhu, Yan, Tersoff,
  and Avouris]{wrinklesElectronic}
Zhu,~W.; Low,~T.; Perebeinos,~V.; Bol,~A.~A.; Zhu,~Y.; Yan,~H.; Tersoff,~J.;
  Avouris,~P. Structure and Electronic Transport in Graphene Wrinkles.
  \emph{Nano Lett.} \textbf{2012}, \emph{12}, 3431--3436\relax
\mciteBstWouldAddEndPuncttrue
\mciteSetBstMidEndSepPunct{\mcitedefaultmidpunct}
{\mcitedefaultendpunct}{\mcitedefaultseppunct}\relax
\EndOfBibitem
\bibitem[Wu \latin{et~al.}(1986)Wu, Hawrylak, Eliasson, Quinn, and
  Fetter]{WU1986795}
Wu,~J.-W.; Hawrylak,~P.; Eliasson,~G.; Quinn,~J.; Fetter,~A. Magnetoplasma
  surface waves on the lateral surface of a semiconductor superlattice.
  \emph{Sol. State Commun.} \textbf{1986}, \emph{58}, 795--798\relax
\mciteBstWouldAddEndPuncttrue
\mciteSetBstMidEndSepPunct{\mcitedefaultmidpunct}
{\mcitedefaultendpunct}{\mcitedefaultseppunct}\relax
\EndOfBibitem
\bibitem[Wang \latin{et~al.}(2011)Wang, Apell, and Kinaret]{PRB84.085423}
Wang,~W.; Apell,~P.; Kinaret,~J. Edge plasmons in graphene nanostructures.
  \emph{Phys. Rev. B} \textbf{2011}, \emph{84}, 085423\relax
\mciteBstWouldAddEndPuncttrue
\mciteSetBstMidEndSepPunct{\mcitedefaultmidpunct}
{\mcitedefaultendpunct}{\mcitedefaultseppunct}\relax
\EndOfBibitem
\bibitem[Abramowitz and Stegun(1965)Abramowitz, and Stegun]{AS}
Abramowitz,~M.; Stegun,~I.~A. \emph{Handbook of Mathematical Functions: with
  Formulas, Graphs, and Mathematical Tables}; Dover: New York, 1965\relax
\mciteBstWouldAddEndPuncttrue
\mciteSetBstMidEndSepPunct{\mcitedefaultmidpunct}
{\mcitedefaultendpunct}{\mcitedefaultseppunct}\relax
\EndOfBibitem
\bibitem[Harrison(2010)]{harrison}
Harrison,~P. \emph{Quantum Wells, Wires and Dots}, 3rd ed.; Wiley, 2010\relax
\mciteBstWouldAddEndPuncttrue
\mciteSetBstMidEndSepPunct{\mcitedefaultmidpunct}
{\mcitedefaultendpunct}{\mcitedefaultseppunct}\relax
\EndOfBibitem
\bibitem[Nielsen \latin{et~al.}(2008)Nielsen, Fernandez-Cuesta, Boltasseva,
  Volkov, Bozhevolnyi, Klukowska, and Kristensen]{Nielsen:2008}
Nielsen,~R.~B.; Fernandez-Cuesta,~I.; Boltasseva,~A.; Volkov,~V.~S.;
  Bozhevolnyi,~S.~I.; Klukowska,~A.; Kristensen,~A. Channel plasmon polariton
  propagation in nanoimprinted V-groove waveguides. \emph{Opt. Lett.}
  \textbf{2008}, \emph{33}, 2800--2802\relax
\mciteBstWouldAddEndPuncttrue
\mciteSetBstMidEndSepPunct{\mcitedefaultmidpunct}
{\mcitedefaultendpunct}{\mcitedefaultseppunct}\relax
\EndOfBibitem
\bibitem[Toscano \latin{et~al.}(2013)Toscano, Raza, Yan, Jeppesen, Xiao, Wubs,
  Jauho, Bozhevolnyi, and Mortensen]{Toscano:13}
Toscano,~G.; Raza,~S.; Yan,~W.; Jeppesen,~C.; Xiao,~S.; Wubs,~M.; Jauho,~A.-P.;
  Bozhevolnyi,~S.~I.; Mortensen,~N.~A. Nonlocal response in plasmonic
  waveguiding with extreme light confinement. \emph{Nanophotonics}
  \textbf{2013}, \emph{2}, 161--166\relax
\mciteBstWouldAddEndPuncttrue
\mciteSetBstMidEndSepPunct{\mcitedefaultmidpunct}
{\mcitedefaultendpunct}{\mcitedefaultseppunct}\relax
\EndOfBibitem
\bibitem[de~Juan \latin{et~al.}(2013)de~Juan, Ma\~nes, and
  Vozmediano]{Manes:2013}
de~Juan,~F.; Ma\~nes,~J.~L.; Vozmediano,~M. A.~H. Gauge fields from strain in
  graphene. \emph{Phys. Rev. B} \textbf{2013}, \emph{87}, 165131\relax
\mciteBstWouldAddEndPuncttrue
\mciteSetBstMidEndSepPunct{\mcitedefaultmidpunct}
{\mcitedefaultendpunct}{\mcitedefaultseppunct}\relax
\EndOfBibitem
\bibitem[Amorim \latin{et~al.}(2016)Amorim, Cortijo, de~Juan, Grushin, Guinea,
  Gutiérrez-Rubio, Ochoa, Parente, Roldán, San-Jose, Schiefele, Sturla, and
  Vozmediano]{Bruno:2016}
Amorim,~B.; Cortijo,~A.; de~Juan,~F.; Grushin,~A.; Guinea,~F.;
  Gutiérrez-Rubio,~A.; Ochoa,~H.; Parente,~V.; Roldán,~R.; San-Jose,~P.;
  Schiefele,~J.; Sturla,~M.; Vozmediano,~M. Novel effects of strains in
  graphene and other two dimensional materials. \emph{Physics Reports}
  \textbf{2016}, \emph{617}, 1 -- 54, Novel effects of strains in graphene and
  other two dimensional materials\relax
\mciteBstWouldAddEndPuncttrue
\mciteSetBstMidEndSepPunct{\mcitedefaultmidpunct}
{\mcitedefaultendpunct}{\mcitedefaultseppunct}\relax
\EndOfBibitem
\end{mcitethebibliography}

\end{document}